\documentclass[9pt,twocolumn,twoside]{pnas-new}
\templatetype{pnasresearcharticle} 
\usepackage{cleveref}
\usepackage{siunitx}
\usepackage{wasysym}

\newcommand{\pd}[0]{\partial}

\newcommand{\eps}[0]{\varepsilon}

\newcommand{\Bo}[0]{\operatorname{Bo}}
\newcommand{\Ca}[0]{\operatorname{Ca}}
\newcommand{\We}[0]{\operatorname{We}}

\title{Fluid dynamics alters liquid-liquid phase separation in confined aqueous two-phase systems}

\author[a,b]{Eric W. Hester}
\author[a,b]{Sean P. Carney} 
\author[c]{Vishwesh Shah} 
\author[c]{Alyssa Arnheim}
\author[c]{Bena Patel}
\author[b,c,d]{Dino Di Carlo} 
\author[a,b,d,1]{Andrea L. Bertozzi}

\affil[a]{Department of Mathematics, University of California Los Angeles, 520 Portola Plaza, Los Angeles, 90095, CA, USA.}
\affil[b]{California NanoSystems Institute,  University of California Los Angeles, Los Angeles, 90095, CA, USA.}
\affil[c]{Department of Bioengineering, University of California Los Angeles, Los Angeles, 90095, CA, USA.}
\affil[d]{Department of Mechanical and Aerospace Engineering and California NanoSystems Institute, University of California Los Angeles, Los Angeles, 90095, CA, USA.}

\leadauthor{Hester}

\authorcontributions{
EWH: Designed research, performed research, development of mathematics and code, analysis of results, wrote the paper.
SPC: Designed research, performed research, development of mathematics and code, critical revision of manuscript.
VS: Performed laboratory experiments, wrote up experimental methods.
AA: Performed laboratory experiments.
BP: Performed laboratory experiments.
DD: Designed research, critical revision of manuscript.
ALB: Designed research, critical revision of manuscript.}
\authordeclaration{The authors declare that we have no competing interests.}
\correspondingauthor{\textsuperscript{1}To whom correspondence should be addressed. E-mail: bertozzi@ucla.edu}

\keywords{Fluid Dynamics $|$ Liquid-Liquid Phase Separation $|$ Aqueous Two-Phase Systems} 

\begin{abstract}
Liquid-liquid phase separation is key to understanding aqueous two-phase systems (ATPS) arising throughout cell biology, medical science, and the pharmaceutical industry.
Controlling the detailed morphology of phase-separating compound droplets leads to new technologies for efficient single-cell analysis, targeted drug delivery, and effective cell scaffolds for wound healing.
We present a computational model of liquid-liquid phase separation relevant to recent laboratory experiments with gelatin-polyethylene glycol mixtures. 
We include buoyancy and surface-tension-driven finite viscosity fluid dynamics with thermally induced phase separation. 
We show that the fluid dynamics greatly alters the evolution and equilibria of the phase separation problem.
Notably, buoyancy plays a critical role in driving the ATPS to energy-minimizing crescent-shaped morphologies and shear flows can generate a tenfold speedup in particle formation.
Neglecting fluid dynamics produces incorrect minimum-energy droplet shapes.
The model allows for optimization of current manufacturing procedures for structured microparticles and improves understanding of ATPS evolution in confined and flowing settings important in biology and biotechnology.
\end{abstract}

\significancestatement{Liquid-liquid phase separation (LLPS) at the microscale creates complex multi-phase droplets relevant to medical science, pharmaceuticals, and cell biology.
We show that fluid flow is key to the evolution and equilibria of such droplets.
Cahn-Hilliard models that neglect fluid dynamics fail to attain minimum-energy equilibria observed in laboratory experiments.
In contrast, models incorporating fluid dynamics reproduce experiments and illustrate how external forces drive internal flows that can accelerate droplet formation.
These results suggest that fluid stresses play an important role in LLPS-driven generation of biomolecular condensates,
and point the way to methods for efficient and robust optimization of droplet-based microparticle manufacture.
}

\dates{This manuscript was compiled on \today}
\doi{\url{www.pnas.org/cgi/doi/10.1073/pnas.XXXXXXXXXX}}

\begin{document}

\maketitle
\thispagestyle{firststyle}
\ifthenelse{\boolean{shortarticle}}{\ifthenelse{\boolean{singlecolumn}}{\abscontentformatted}{\abscontent}}{}

\dropcap{L}iquid-liquid phase separation (LLPS) powers versatile techniques for creating complex microstructures useful throughout the medical, agricultural, and pharmaceutical industries \cite{XiaoReviewPreparationApplication2014,
MaCellInspiredAllAqueousMicrofluidics2020}.
LLPS also explains membraneless organelles (biocondensates) arising in cell biology \cite{HymanLiquidLiquidPhaseSeparation2014,
BananiBiomolecularCondensatesOrganizers2017,
BerryPhysicalPrinciplesIntracellular2018} as well as several indicators and causes of cell dysfunction \cite{ShinLiquidPhaseCondensation2017,
AlbertiLiquidLiquidPhase2019}.
Recent work has sought to design morphologies of many-component phase-separating liquids by controlling surface energies and volume fractions of each phase \cite{MaoPhaseBehaviorMorphology2019,MaoDesigningMorphologySeparated2020,ShrinivasPhaseSeparationFluids2021,ZhangPhaseFieldModelingMultiple2021}.
Here, we demonstrate the critical role of fluid dynamics in liquid-liquid phase separation.
Specifically, we present a combined experimental, theoretical, and numerical investigation of an aqueous two-phase system (ATPS) consisting of a spherical drop of gelatin-polyethylene glycol (PEG) polymer solution suspended in a surrounding continuous phase that undergoes temperature-induced phase separation at \SI{4}{\celsius} (similar to \cite{YanagisawaPhaseSeparationBinary2014}).
Morphology design for this system also has direct relevance to high-throughput manufacture of microparticles \cite{LeeScalableFabricationUse2022,
WangCountingEnzymaticallyAmplified2021,
deRutteSuspendableHydrogelNanovials2022} used for scalable single-cell analysis \cite{ThebergeMicrodropletsMicrofluidicsEvolving2010,
ZarzarDynamicallyReconfigurableComplex2015,
LiMicrofluidicFabricationMicroparticles2018,
CaiAnisotropicMicroparticlesMicrofluidics2021,
KimDropletMicrofluidicsProducing2014,
ShangEmergingDropletMicrofluidics2017}, 
where LLPS obviates the need for complex flow-focussing microfluidics devices when constructing Janus microparticles \cite{
MaFabricationMicrogelParticles2012,
WangHoleShellMicroparticles2013,
WatanabeMicrofluidicFormationHydrogel2019,
LiuSelfOrientingHydrogelMicroBuckets2019,
ChenMicrofluidicGeneratedBiopolymerMicroparticles2021}.

\begin{figure*}[hbt]
    \includegraphics[width=\linewidth]{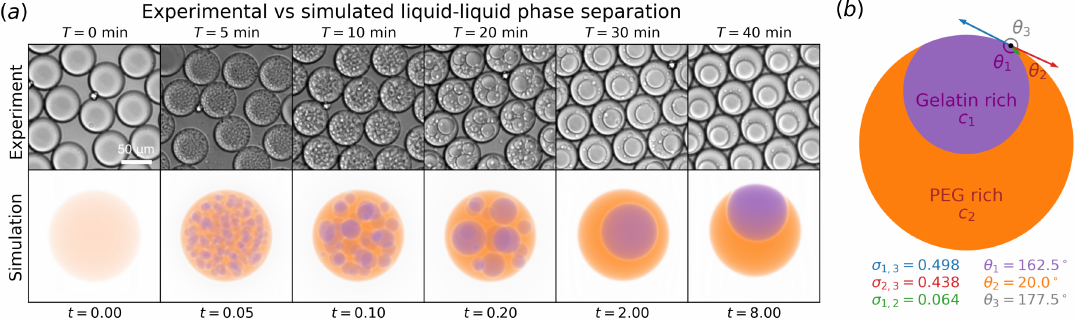}
    \caption{
    ($a$): The first row shows microscopic images of thermally induced phase separation in a gelatin-PEG mixture at \SI{4}{\celsius} with increasing time.
    Each droplet measures 50 microns in diameter.
    Experimental details are provided in appendix \ref{sec:methods}.
    The second row plots a volume rendering of a three dimensional simulation of liquid-liquid phase separation using the Cahn-Hilliard-Stokes-Boussinesq model developed in section 1 \ref{sec:CHSB} (corresponding to model CHSB $x$ in \cref{fig:chsb-3D}).
    Simulation parameters are detailed in \cref{tab:dimensional-parameters,tab:nondimensional-parameters,tab:numerical-parameters}.
    The simulation time $t$ is nondimensional, and must be rescaled to correspond to the physical time $T$ of the experimental snapshots, as we use an artificially large interfacial thickness parameter $\eps$ (detailed in the supplementary material).
    The patterns of phase separation evolving over time are in qualitative agreement.
    ($b$): Schematic of minimal-energy crescent-shaped particle at equilibrium for given surface tensions (red, green, blue) and contact angles (purple, orange, gray) of gelatin-rich ($c_1$), PEG-rich ($c_2$), and surrounding oil ($c_3$) phases.}
    \label{fig:experiment-vs-sim-big}

\end{figure*}

In section \ref{sec:theory} we introduce a hierarchical suite of models of LLPS. 
We begin with surface-energy minimization (section 1 \ref{sec:surface-energy}), 
add spinodal decomposition with a ternary extension of Cahn-Hilliard/Model B \cite{HohenbergTheoryDynamicCritical1977}, 
and finally incorporate surface tension- and buoyancy-driven incompressible viscous fluid dynamics with the Cahn-Hilliard-Stokes-Boussinesq model, an extension of Model H \cite{HohenbergTheoryDynamicCritical1977} (section 1 \ref{sec:CHSB}).
In section \ref{sec:results} we illustrate the predictions of the surface-energy minimizing model (section 2 \ref{sec:minimal-energy-results}), demonstrate the failure of Cahn-Hilliard to reproduce experiments (section 2 \ref{sec:ch-results}), and then explore the effects of fluid dynamics on liquid-liquid phase separation in section 2 \ref{sec:chsb-results}. 
In particular, we show that:
\begin{enumerate}
\item Cahn-Hilliard/Model B dynamics starting from mixed initial conditions evolves to a core-shell morphology, rather than energy minimizing crescents.
\item Cahn-Hilliard-Stokes-Boussinesq/Model H dynamics incorporating fluid forces evolves to the experimentally-observed minimial-energy crescent shapes (\cref{fig:experiment-vs-sim-big}).
\item Shear-induced recirculation, e.g. arising from pressure-driven channel flows, can drive a tenfold acceleration in crescent formation, speeding microparticle manufacture.
\end{enumerate}
We finally conclude with future directions for modeling fluid dynamics in liquid-liquid phase separation in \cref{sec:conclusion}.
We expect our model to help optimize the speed of microparticle manufacture \cite{LeeScalableFabricationUse2022,
deRutteSuspendableHydrogelNanovials2022},
as well as the design of new classes of structured microparticles for drug delivery and tissue engineering.
More broadly, our work bolsters the emerging importance of fluid dynamics \cite{PaulsenCoalescenceBubblesDrops2014,
ShimizuNovelCoarseningMechanism2015} in industrial \cite{SteinbacherPolymerChemistryFlow2006,
ZhouPhaseFieldSimulations2006,
LiPhasefieldSimulationThermally2012,
TreeMasstransferDrivenSpinodal2019,
WangPhasefieldStudyPolymerizationinduced2020}, biological \cite{
BerryRNATranscriptionModulates2015,
BerryPhysicalPrinciplesIntracellular2018}, 
and mathematical \cite{MaoPhaseBehaviorMorphology2019,MaoDesigningMorphologySeparated2020,ShrinivasPhaseSeparationFluids2021} LLPS problems.

\section{Theory}
\label{sec:theory}

\subsection{Surface energy minimization}
\label{sec:surface-energy}
The simplest model of ternary fluids predicts each phase will arrange itself to minimize the total interfacial energy.
This isoperimetric problem leads to interfaces that are either flat, or a portion of a sphere \cite{BostwickStabilityConstrainedCapillary2015}.
At triple contact points, the angle spanned by each phase satisfies a force balance given by the Young condition
	\begin{align}
	\frac{\sigma_{1,2}}{\sin \theta_3} = 
	\frac{\sigma_{1,3}}{\sin \theta_2} &= 
	\frac{\sigma_{2,3}}{\sin \theta_1}, &
	\theta_1 + \theta_2 + \theta_3 &= 2 \pi,
	\end{align}
where $\sigma_{i,j}$ is the surface tension across the $i,j$ interface.
Triple contact points are unstable if any surface tension $\sigma_{i,j}$ dominates the sum of the remaining two, or equivalently if any wetting parameter $\chi_{i} = \sigma_{i,i+1} + \sigma_{i,i+2} - \sigma_{i+1,i+2}$ is negative (with indices modulo $\{1,2,3\}$).
The supplementary material solves this minimization problem to give explicit formulae for minimal energy shapes in ATPS droplets as a function of volume ratio and surface tensions.

\subsection{Nonequilibrium models: Cahn-Hilliard-Stokes-Boussinesq}
\label{sec:CHSB}
The surface energy minimization model only predicts equilibrium shapes.
But droplet formation is inherently nonequilibrium, involving both thermally induced phase separation and fluid flow.
To model dynamic phase separation from an initially mixed model we begin with a ternary Cahn-Hilliard model from \cite{DongMultiphaseFlowsImmiscible2018}, which generalizes several earlier advective Cahn-Hilliard models \cite{AndersonDiffuseInterfaceMethodsFluid1998,JacqminCalculationTwoPhaseNavier1999,KimPhaseFieldModeling2005,BoyerStudyThreeComponent2006,BoyerHierarchyConsistentNcomponent2014}.
After transforming to dimensionless quantities (see supplementary material), we neglect inertia and apply the Boussinesq approximation (ignoring density variation aside from the buoyancy term) at constant viscosity to derive the Cahn-Hilliard-Stokes-Boussinesq (CHSB) equations
 	\begin{align}
	\pd_t c_i + u \cdot \nabla c_i - \sum_{j=1}^3 m_{ij}\nabla^2 \mu_j &= 0,\\
	\nabla p - \Ca \nabla^2 u - \Bo \rho \, \hat{g} &= \frac{3}{\sqrt{2}}\frac{1}{\eps} \sum_{i=1}^3\mu_i \nabla c_i,\\
	\nabla \cdot u &= 0,
	\end{align}
for the concentrations $c_i$ (constrained by $\sum_{i=1}^3 c_i = 1$), the fluid velocity $u$ and pressure $p$.
The chemical potentials $\mu_i$, perturbation density $\rho$, and mobility tensor $m_{ij}$ are given by 
	\begin{align}
	\mu_i &= 2\chi_i c_i(1-c_i)(1-2c_i) + \eps^2 \sum_{j=1}^3 \sigma_{i,j}\Delta c_j,\\
    \rho &= \sum_{i=1}^3 \rho_i c_i, \qquad \qquad 
	m_{ij} = \begin{cases}
		2 & i = j,\\
		-1 & i \neq j.
	\end{cases}
	\end{align}
The key dimensionless parameters are the interface thickness $\eps$, the capillary number $\Ca$, Bond number $\Bo$ and Weber number $\We$, which compare viscosity, buoyancy, and inertia with surface tension forces, respectively:
	\begin{align}
	\eps &= \frac{\bar{\eps}}{L}, &
	\Ca &= \frac{\rho_0 \nu_0 L}{\sigma_0 T}, &
	\Bo &= \frac{\Delta \rho \, g L^2}{\sigma_0}, &
	\We &= \frac{\rho_0 L^3}{\sigma_0 T^2},
	\end{align}
where $\bar{\eps}$ is the dimensional interface thickness between phases, $L$ a characteristic droplet length scale, $\rho_0$ the average density, $\Delta \rho$ the density variation scale, $\nu_0$ the average viscosity, $\sigma_0$ the sum of dimensional surface tensions, $g$ the acceleration due to gravity, $\hat{g}$ the gravitational unit vector, and $T$ the characteristic separation time scale.
Droplet formation dynamics depends on the relative sizes of these dimensionless numbers.
We note that the equilibrium of the Cahn-Hilliard model reduces to the surface-energy minimization model in the limit of vanishing interface thickness ${\eps} \to 0$ \cite{AlikakosConvergenceCahnHilliardEquation1994}.
Our goal is to investigate the role of buoyancy forces in the CHSB equations, and to contrast this behavior with the ternary Cahn-Hilliard (CH) model, which simply sets the fluid velocity to zero $u=0$.

\section{Results and discussion}
\label{sec:results}
\subsection{Minimal energy configurations}
\label{sec:minimal-energy-results}
\begin{figure*}
	\centering
	\includegraphics[width=\linewidth]{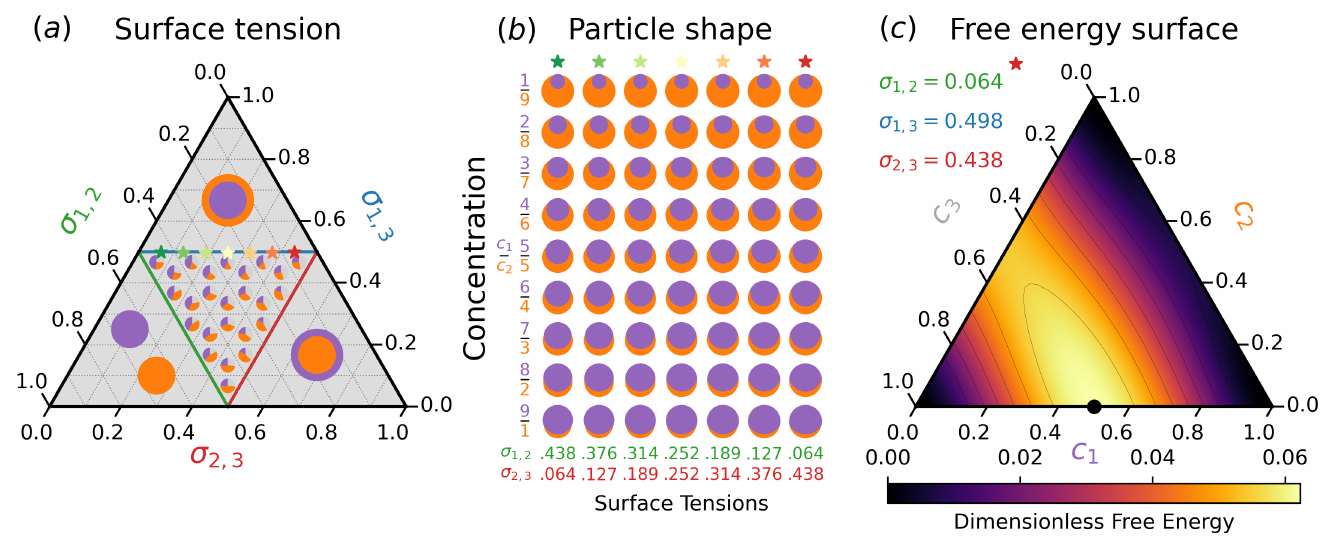}
	\caption{$(a)$: Possible contact angles as a function of relative surface tensions $\sigma_{i,j}$.
	Wetting occurs if any surface tension dominates the other two (the top, left, and right subtriangles of ternary diagram).
	$(b)$:	Example minimal energy configurations for varying surface tensions $\sigma_{i,j}$ (horizontal axis) and relative concentrations $c_1/c_2$ (vertical axis).
	Surface tensions are chosen so that $\theta_2 = 20^\circ$.
	The surface tensions of each column are indicated in the ternary plot with stars shifting from green ($\sigma_{1,2} \approx 0.5$) to red ($\sigma_{2,3} \approx 0.5$).
    $(c)$: The free energy used in simulations of experiments. The maximum is located at the black circle. The energy is more sensitive to $c_3$ than $c_1$ or $c_2$.}
	\label{fig:surface-tension-shape-free-energy-diagram}
\end{figure*}

Four possible minimal energy regimes arise depending on the relative surface tensions.
Scaling dimensionless surface tensions such that $\sigma_{1,2}+\sigma_{1,3}+\sigma_{2,3} = 1$ allows convenient representation on a ternary diagram.
We summarize these regimes in \cref{fig:surface-tension-shape-free-energy-diagram} ($a,b$).
There are three separate wetting regimes, in which either $\sigma_{1,2}, \sigma_{1,3}$ or $\sigma_{2,3}$ dominates the sum of the other two tensions.
These correspond to separated drops, 1-in-2 shells, or 2-in-1 shells (the left, top, and right regions of \cref{fig:surface-tension-shape-free-energy-diagram} ($a$) respectively).
Between these regions we can achieve stable triple points and non-trivial droplet shapes (\cref{fig:surface-tension-shape-free-energy-diagram} ($b$)).
These crescent shapes vary with two independent surface tension parameters, as well as the relative concentration of $c_1$ to $c_2$.
Configurations close to the edges of the inner region give a single small angle (e.g. large $\sigma_{1,3}$ gives small $\theta_2$).
Explicit formulae for these shapes are provided in the supplementary material.
We note that in the case of 1-in-2 or 2-in-1 shells the energy minimizer is not unique.
That is, the total energy does not depend on the location of the inner droplet (provided it does not intersect the outer shell).
However, a near-crescent configuration is neither topologically or energetically equivalent to the actual crescents observed in experiments.

We now focus on the experimental parameter regimes in \cref{tab:dimensional-parameters,tab:nondimensional-parameters}.
Our laboratory experiments exhibit crescents with a contact angle $\theta_2$ of approximately $20^\circ$ \cite{LeeScalableFabricationUse2022}.
This constrains the possible surface tensions to lie close to the edge $\sigma_{1,3} = 0.5$.
We illustrate seven different possible contact angles in \cref{fig:surface-tension-shape-free-energy-diagram} ($b$), that vary from small $\sigma_{1,2}$ (where $\theta_3 \approx 180^\circ$) to small $\sigma_{2,3}$ (where $\theta_1 \approx 180^\circ$).
Physically, we expect a lower surface tension between the aqueous polymer mixtures than between each mixture with the surrounding third phase.
As such, we focus on the case with largest $\theta_3$ (corresponding to ${\sigma_{1,2} = 0.064, \sigma_{1,3} = 0.498, \sigma_{2,3}=0.438}$) for which we plot the free energy surface in (\cref{fig:surface-tension-shape-free-energy-diagram} $(c)$).
The range of surface tensions leading to small contact angles is exceedingly narrow.
Surface tensions must therefore be precisely tuned to achieve crescent particles that surround a significant volume.
Finally, while this model is useful for understanding possible equilibrium shapes, and inferring necessary surface tensions for a desired shape, it does not capture the dynamics of liquid-liquid phase separation (LLPS).

\subsection{The failure of ternary Cahn-Hilliard dynamics}
\label{sec:ch-results}
\begin{figure}
	\centering
	\includegraphics[width=\linewidth]{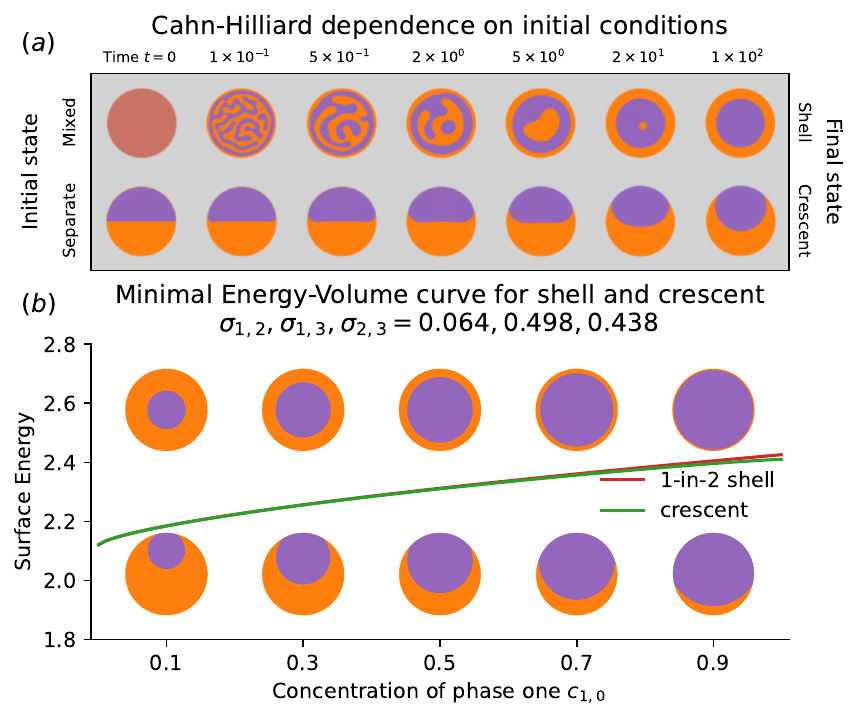}
	\caption{
	$(a)$: Evolution of ternary Cahn-Hilliard model from different initial conditions (mixed vs separate). 
	For equal concentration ratio $c_1:c_2 = 0.5$, mixed initial conditions lead to shells, while initially separated drops form minimal-energy crescents.
	$(b)$ Energy-volume curve for the shell (red) and crescent (green) configurations of figure ($a$) with surface tensions of \cref{fig:surface-tension-shape-free-energy-diagram} ($c$).
	}
	\label{fig:ch-evolution-shape-energy-comparison}
\end{figure}

A sensible start for a model of dynamic LLPS is a three-phase extension of the canonical Cahn-Hilliard equation \cite{CahnFreeEnergyNonuniform1958,HohenbergTheoryDynamicCritical1977}.
The Cahn-Hilliard model tends to a Mullins-Sekerka problem as the interface thickness $\eps$ tends to zero \cite{AlikakosConvergenceCahnHilliardEquation1994}.
Equilibria of the former should thus recover the same energy-minimizing equilbria of the latter -- equivalent to the energy minimizing model of section 1 \ref{sec:surface-energy}. 
However, we show that this model does not always attain the \emph{global} minimal surface energy configuration from section 2 \ref{sec:minimal-energy-results}, and that the final state instead depends on the choice of initial conditions.
In \cref{fig:ch-evolution-shape-energy-comparison} $(a)$ we simulate the CH model for experimental parameters in \cref{tab:dimensional-parameters,tab:nondimensional-parameters} and numerical parameters in \cref{tab:numerical-parameters}, and compare the evolution of a compound droplet with equal $c_1,c_2$ concentrations starting from mixed (top row) and separated (bottom row) initial conditions.
While the initially separated particle evolves to the minimal energy crescent, the initially mixed droplet (corresponding to the experiment in \cref{fig:experiment-vs-sim-big}) evolves to a stable core-shell droplet shape after undergoing spinodal decomposition.
\Cref{fig:ch-evolution-shape-energy-comparison} $(b)$ hints at why the CH model fails to attain the minimal-energy crescent shape when evolving from mixed initial conditions: the small value of $\sigma_{1,2}$ means that the 1-in-2 shell is very close in energy to the crescent configuration.
Similar patterns have been observed in earlier three phase simulations of membrane manufacture in Cartesian geometries \cite{ZhouPhaseFieldSimulations2006}.
These solutions do not evolve to an asymmetric crescent shape from well-mixed initial conditions, in contrast to our experiment.
We must incorporate additional relevant physics. 
An obvious choice is fluid dynamics, either through buoyancy forces, viscous shear instabilities, or other destabilizing phenomena.

\subsection{Evolution simulations}
\label{sec:chsb-results}
\begin{figure*}
	\centering
	\includegraphics[width=\linewidth]{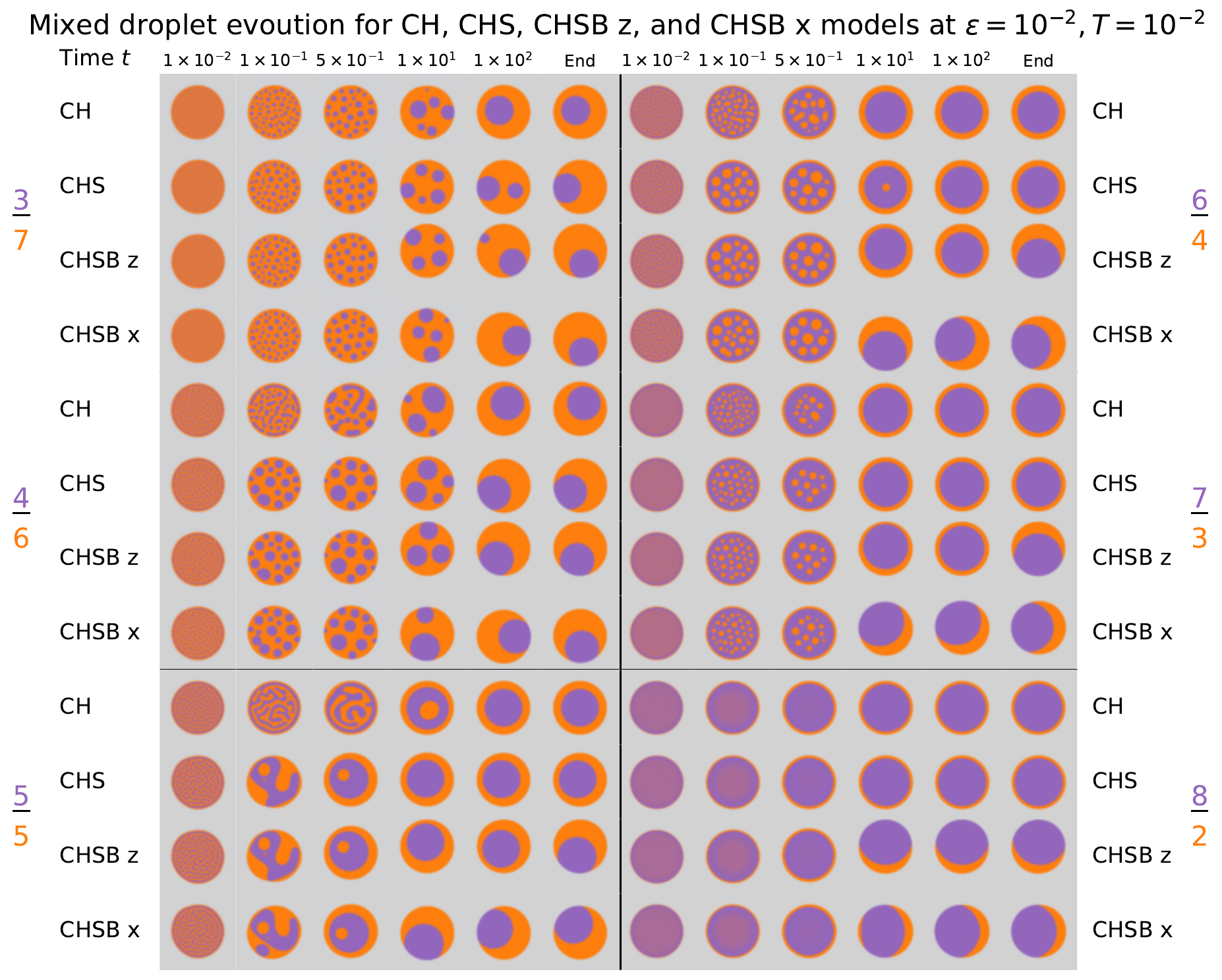}
	\caption{
        Comparison of CH, CHS, and CHSB ($x, z$) models in two dimensions for concentration ratios $c_1 = 0.3,\ldots,0.8$, for six time snapshots, under parameters given in \cref{tab:dimensional-parameters,tab:nondimensional-parameters,tab:numerical-parameters}.
        Several regimes are apparent for different choices of initial concentration in each model.  
        The final shape and time to equilibrium are strongly dependent of the presence and type of perturbations due to asymmetric fluid forces.
        }
	\label{fig:chs-2D-comparison-T-1e-2}
\end{figure*}

Using the parameters inferred from experimental contact angles, we compare four different models:
\begin{enumerate}
	\item (CH): The ternary Cahn-Hilliard model,
	\item (CHS): The Cahn-Hilliard-Stokes model incorporating fluid dynamics but without buoyancy,
	\item (CHSB $z$): The Cahn-Hilliard-Stokes-Boussinesq model including phase separation, fluid flow, and buoyancy forcing normal to a no-slip channel,
	\item (CHSB $x$): The Cahn-Hilliard-Stokes-Boussinesq model with buoyancy forcing parallel to a no-slip channel,
\end{enumerate}
and six different initial concentrations (\cref{fig:chs-2D-comparison-T-1e-2}).

We estimate physical scales in \cref{tab:dimensional-parameters}, from which we derive the nondimensional parameters in \cref{tab:nondimensional-parameters}.
Precise dimensional parameters are difficult to measure.
In particular, the interfacial length scale $\bar{\eps}$ and the mobility (indirectly determined through the time scale $T$) are not directly measurable.
They are instead constrained by computational resources.
Here we choose $\bar{\eps} = \SI{1e-6}{m}$ and $T = \SI{1e-2}{s}$ to balance accuracy (smaller $\bar{\eps}$ and larger $T$) with efficiency (larger $\bar{\eps}$ and smaller $T$).
The supplementary material has additional simulations at varying $\eps$ to validate our choice of model and parameters.

The numerical parameters for the simulations are provided in \cref{tab:numerical-parameters}, with more details provided in supplementary material.
We examine higher resolution simulations in both two dimensional and three-dimensional simulations
to fully capture the role of fluid dynamics.

\begin{table}[hbt]
	\centering
	\begin{tabular}{lcl}
    Name & Symbol & Value\\
    \hline
    Time scale & $T$ & \SI{1e-2}{s}\\
    Length scale & $L$ & \SI{1e-4}{m}\\
    Surface tension scale & $\sigma_0$ & \SI{1e-2}{kg.s^{-2}}\\
    Mass density scale & $\rho_0$ & \SI{1e3}{kg.m^{-3}}\\
    Mass density variation scale & $\Delta \rho_0$ & $\SI{1e3}{kg.m^{-3}}$\\
    Kinematic viscosity & $\nu_0$ & \SI{1e-6}{m^2.s^{-1}}\\
	Gravity & $g$ & \SI{1e2}{m.s^{-2}}\\
	\hline
	\end{tabular}
	\caption{Dimensional parameters for simulations.}
	\label{tab:dimensional-parameters}

	\vspace{1em}	
	\begin{tabular}{lcl}
    Name & Symbol & Value\\
    \hline
    Interface thickness & $\eps$ & \num{1e-2}\\
    Bond number & Bo & \num{1e-3}\\
    Capillary number & Ca & \num{1e-3}\\
    Weber number & We & \num{1e-3}\\
    Surface tensions & $\sigma_{1,2}, \sigma_{1,3},\sigma_{2,3}$ & $0.0636,	0.4983,	0.4381$\\
    Density perturbations & $d\rho_1, d\rho_2,d\rho_3$ & $0.05,-0.01,0.6$\\
    Initial concentrations & $c_{1,0}, c_{2,0}$ & $0.3,0.4,\ldots,0.8$\\
	\hline
	\end{tabular}
	\caption{Nondimensional parameters for simulations.}
	\label{tab:nondimensional-parameters}

	\vspace{1em}	
	\begin{tabular}{ll}
    Quantity & Value\\
    \hline
    Spatial modes $n_x,n_y,n_z$ & $192, 192, 384$\\
    Dealias factor & $3/2$\\
    Time step $dt$ & \num{2e-4} \\
    Time step scheme & SBDF2\\
	\hline
	\end{tabular}
	\caption{Numerical parameters for simulations.}
	\label{tab:numerical-parameters}
\end{table}
 
\begin{figure*}
	\centering
	\includegraphics[width=.9\linewidth]{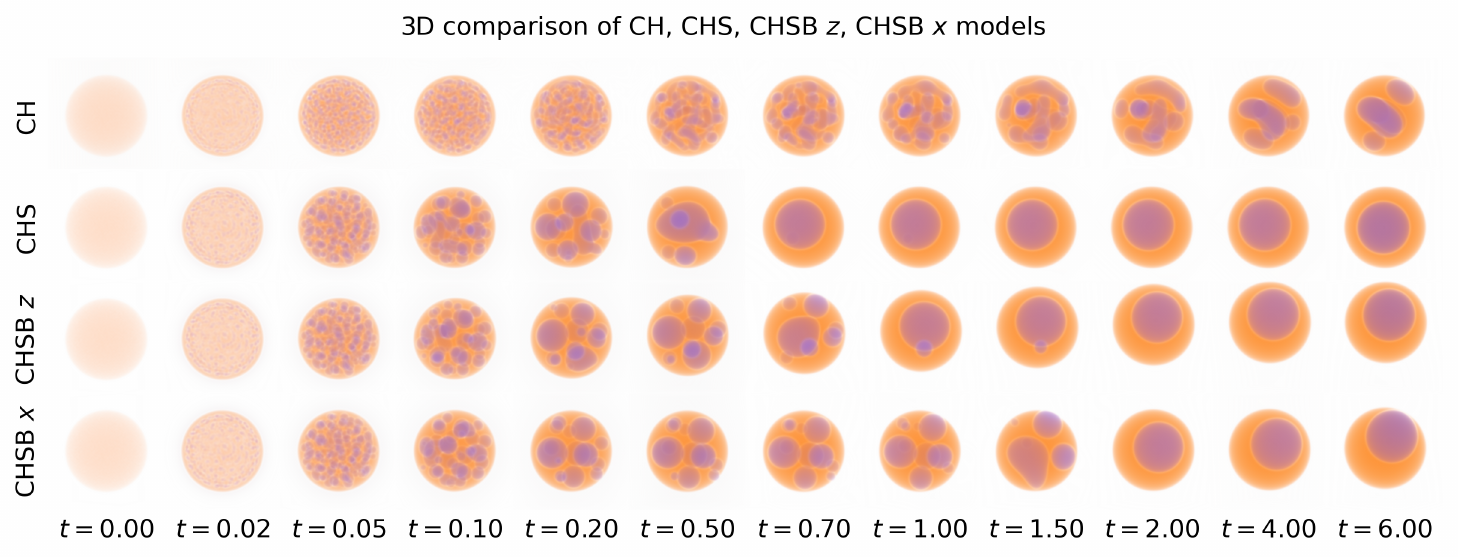}
	\caption{
	Evolution of CHS, CHSB $z$ and CHSB $x$ simulations in three dimensions using parameters in \cref{tab:dimensional-parameters,tab:nondimensional-parameters,tab:numerical-parameters}, for the initial concentration ratio $c_1:c_2 = 0.3:0.7$.
    Overall coarsening times are accelerated compared to two dimensional simulations.
    Only the CHSB $x$ model develops an energy minimizing crescent configuration.
	}
	\label{fig:chsb-3D}
\end{figure*}

\begin{figure}
    \centering
    \includegraphics[width=\linewidth]{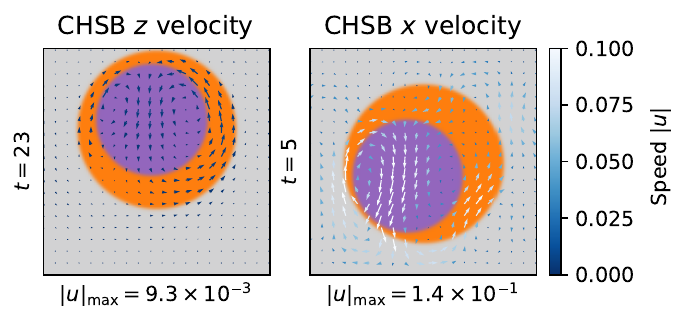}
    \caption{
    Example snapshots of the velocity field for the 2D CHSB $z$ (left) and CHSB $x$ (right) models.
    Both simulations have initial concentration ratio $c_1:c_2 = 0.5$.
    The phases are illustrated in color ($c_1,c_2,c_3$ are purple/orange/gray) and the velocities indicated with arrows (colored by speed).
    The time of each snapshot is provided on the left of each axis, and the maximum speed indicated on the bottom.
    The CHSB $x$ model has been shifted into the droplet frame, and we have subtracted off the $x$-averaged velocity profile to visualize the deviation from Poiseuille-type shear flow.
    The length scales of the velocities in each figure are different, however both are colored according to the same velocity scale (right).
    }
    \label{fig:velocity}
\end{figure}

\begin{figure}
    \centering
    \includegraphics[width=\linewidth]{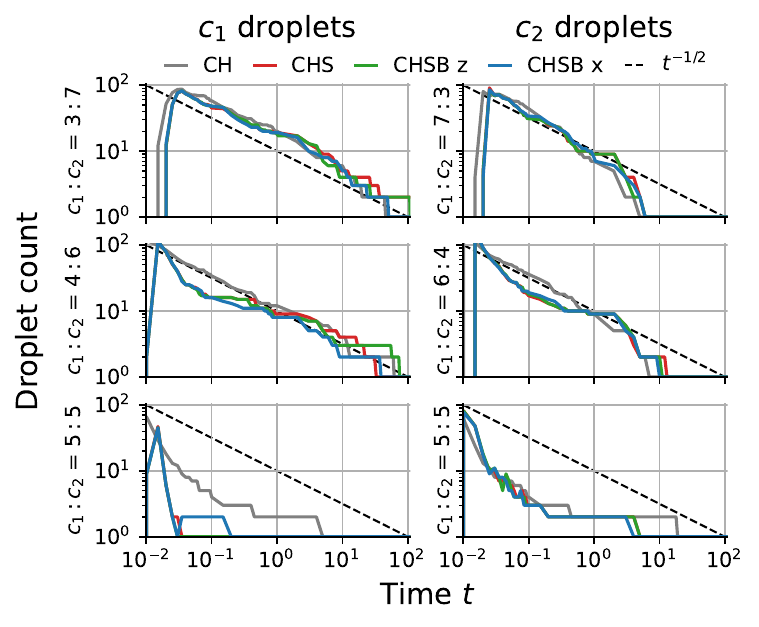}
    \caption{
    Plot of number of $c_1$ droplets (left column) and $c_2$ droplets (right column) over time, for different concentration ratios $c_1:c_2$, for CH (gray), CHS (red), CHSB $z$ (green) and CHSB $x$ models (blue).
    A $t^{-1/2}$ line (dashed black) fits some coalescence regimes well.
    The CHS, CHSB $z$, and CHSB $x$ models show very little difference in particle counts.
    The CH model is similar for concentration ratios $c_1:c_2$ of $3:7$ and $7:3$, but becomes increasingly distinct as the concentration ratio approaches $5:5$.
    Droplet counts are omitted for other choices of droplet and concentration ratio because at most one drop ever forms.
    }
    \label{fig:particle-number-vs-time}
\end{figure}

\begin{figure}
	\centering
	\includegraphics[width=\linewidth]{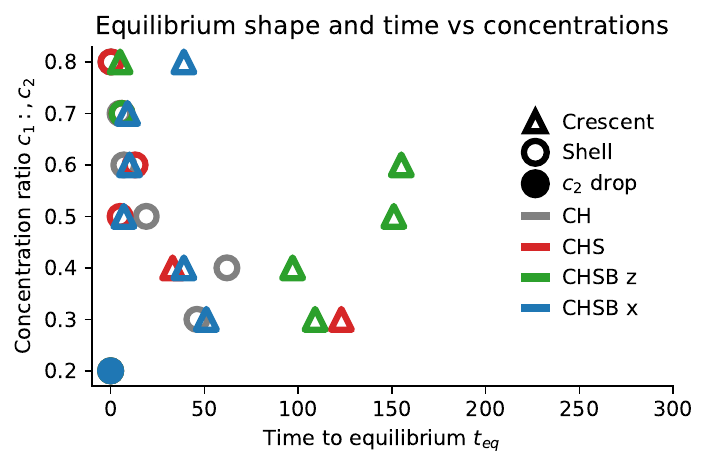}
	\caption{
	Equilibrium shape (Crescent $\triangle$, shell $\fullmoon$, and $c_2$ droplet $\newmoon$) and dimensionless time to equilibrium for CH (gray), CHS (red), CHSB $z$ (green), and CHSB $x$ (blue) models.
        Asymmetric buoyancy forces destabilize shell configurations, causing crescent formation.
        Pressure driven pipe flow causes stronger recirculation, accelerating crescent formation by up to two orders of magnitude. 
        Low $c_1$ concentrations instead diffuse out of the drop, leading to $c_2$-only drops.
	}
	\label{fig:end_times}
\end{figure}

\subsubsection{Concentration dependent coarsening}
The initial coarsening in \cref{fig:chs-2D-comparison-T-1e-2} (also supplementary movie S1) for $t \leq \num{5e-1}$ reveals a strong dependence on initial concentration ratios, moderate influence of fluid dynamics, and weak influence of fluid perturbations due to buoyancy.
At least five regimes can be discerned.
(1) The first, at low initial $c_{1,0}$ concentrations ($c_{1,0} < 0.3$), leads to complete dissipation of $c_1$ throughout the domain (omitted from \cref{fig:chs-2D-comparison-T-1e-2}).
No area forms a critical mass sufficient to nucleate a $c_1$ droplet, and the droplet concentration becomes homogeneous.
(2) For intermediate $c_2$-dominated concentrations $(0.3 \leq c_{1,0} < 0.5)$, the initial separation dynamics generate coarsening $c_1$ droplets embedded in a connected $c_2$ phase. Some $c_1$ droplets form sufficiently close to the boundary to create a triple point.
(3) For balanced initial conditions $c_{1,0} = 0.5$, the initial coarsening generates a mixture of two continuous phases, rather than one phase of small separated droplets.
This leads to much faster coarsening;
the CHS and CHSB systems settle to a 1-in-2 shell by $T = 0.5$, whereas multiple droplets remain at $T=10$ for lower initial $c_{1,0}$ concentrations.
We also observe the most pronounced differences between the CH model and the models incorporating fluid dynamics.
By $t=0.1$ the CHS and CHSB systems have coarse and rounded features similar to experiments, while the CH model still demonstrates labyrinthine patterns of large aspect ratio. 
The evolution of the CH model is also much delayed compared to models incorporating fluid dynamics, as has been observed in other two-phase fluid studies 
\cite{TjahjadiSatelliteSubsatelliteFormation1992,StoneDynamicsDropDeformation1994}.
(4) For intermediate $c_1$-dominated concentrations $(0.5 < c_{1,0} \leq 0.7)$, 
the initial evolution of the system leads to coarsening $c_2$ droplets within a continuous $c_1$ phase, which is itself enveloped by a thin $c_2$ shell (unlike the small $c_{1,0}$ case).
The discrepancy between the CH and remaining models is also much reduced, almost disappearing for concentration ratio $0.8$.
(5) Finally, for sufficiently high concentrations of $c_1$, complete dissipation of $c_2$ droplets occurs and a thin $c_2$ skin layer forms surrounding a homogeneous $c_1$ droplet.

We quantify these coarsening observations in \cref{fig:particle-number-vs-time}, where we plot the number of $c_1$ droplets over time for concentration ratios $c_{1,0} = 0.3,0.4,0.5$, and the number of $c_2$ droplets over time for $c_{1,0} = 0.5,0.6,0.7$, for each of the four models.
Very little difference is apparent between the four models in droplet counts.
For low $c_{1,0} < 0.5$ or high $c_{1,0} > 0.5$ concentrations the particle counts are reasonably approximated by a $t^{-1/2}$ coarsening law.
This corresponds to an area per droplet scaling with $t^{1/2}$, and a droplet length scale proportional to $t^{1/4}$.
This agreement worsens near equal concentration ratios, where the system quickly evolves to two interpenetrating phases.
The rate $t^{1/4}$ is slower than the standard diffusion-limited coarsening scaling exponent $n=1/3$ \cite{BerryPhysicalPrinciplesIntracellular2018}.

While coarsening does not immediately demonstrate the importance of fluid dynamics for LLPS, this changes when considering the long time behavior of each system.
Over longer time scales ($t \approx 10^2$) the system settles into a stable equilibrium given by either a shell or crescent configuration.
The long time behavior now differs between the CH, CHS, CHSB $x$, and CHSB $z$ models.
We quantify the final shape of each simulation and the time it took to reach this shape in \cref{fig:end_times}.
As observed in section 2 \ref{sec:ch-results}, the CH model consistently selects a 1-in-2 core-shell morphology, despite the temporary formation of triple points for low initial concentration ratios ($c_{1,0} < 0.5$).
The CHS model similarly fails to attain the minimal energy crescent for $c_{1,0} > 0.3$, though it does form a crescent at $c_{1,0} = 0.3$.
It is somewhat unintuitive that reducing the initial $c_{2,0}$ concentration has improved the robustness of the $c_2$ shell.
A detailed perturbation analysis may resolve this finding, {but is out of scope for this work}.

In contrast, the CHSB models are sufficiently perturbed by buoyancy and shear flows to attain the minimal energy crescent.
We observe a tenfold difference in time to equilibrium shape between the CHSB $z$ and CHSB $x$ models.
The large fluid flows generated by pressure gradients in the CHSB $x$ model destabilize the 1-in-2 droplet configuration, leading to early formation of a crescent at $t = 10$.
The CHSB $z$ model takes until at least $t=100$ to achieve the minimal energy crescent for intermediate concentration ratios $0.2 < c_{1,0} < 0.8$.

We provide an example snapshot of the velocity field in \cref{fig:velocity}.
We note that the CHSB $x$ figure actually plots the \emph{deviation} of the velocity from the $x$-averaged velocity profile.
This is because an $\mathcal{O}(1)$ Poiseuille-type flow is set up by the pressure gradient, obscuring the smaller relative motion of the fluid within the droplet.
The overall magnitude of the recirculation velocity within the CHSB $z$ droplet is an order of magnitude weaker than that in the CHSB $x$ droplet.
Once the CHSB $z$ droplet has reached the top of the constraining box, only much weaker relative buoyancies between the $c_1$ and $c_2$ phase can drive the flow.
This contrasts with the large shear-induced recirculation within the CHSB $x$ droplet.
This difference of flows within the droplets is the origin of the reduced interval to droplet formation in the CHSB $x$ model.
This suggests that crescent particle manufacture may be accelerated by pressure driven shear flow through small confined channels.

In summary, the initial phase separation may dissipate $c_1$, generate droplets of $c_1$ or $c_2$, or form $c_2$ shells, depending on the initial concentration ratio $c_{1,0}$.
The long time behavior will then either lead to a uniform $c_2$ droplet, a 1-in-2 shell, or a minimal energy crescent.
The time to equilibrium depends on both the initial concentrations, as well as the strength of the fluid perturbations.
If particles move due to pressure gradients, leading to shear flows, one observes recirculating flows within the droplet.
This recirculating flow accelerates transport of the first phase to the droplet exterior, aiding crescent formation.
If instead particles are constrained by a bounding box, the buoyancy induced flows are weaker, causing much longer coalescence times.
Without density variations between phase 1 and 2, the shell configuration remains stable, preventing crescent formation.
We provide a side by side video of these simulations in the supplementary material.

\subsubsection{Three dimensional drop evolution}
We present three-dimensional simulations with initial droplet concentration ratio $c_{1,0} :c_{2,0} = 0.3:0.7$ in \cref{fig:chsb-3D} (and supplementary movie S2).
Our aim is to delineate agreement and disagreement between the three-dimensional simulations, the two-dimensional simulations, and experimental data.
For the three-dimensional models, we see agreement at early times $t < 0.5$.
Differences in time to form a single $c_1$ bubble occur between $t = 0.5$ to $t = 2$, where stronger fluid perturbations temporarily inhibit coarsening (the CHS model coarsens to a single droplet by time $t=.7$, in comparison to $t=1.0$ for the CHSB $z$ model and $t=2.0$ for the CHSB $x$ model).
The CH and CHS models tend to a core-in-shell configuration whereas the CHSB $x$ forms a crescent. 
CHSB $z$ should also reach an equilibrium crescent shape, though over a longer time period than our simulation.
We provide a side-by-side video of the three dimensional simulations in the supplementary material.

The patterns of shell vs crescent formation in two and three dimensions are in agreement --- stronger recirculating flows in the CHSB $x$ model accelerate crescent formation compared to the CHSB $z$ model, and the CH and CHS models settle into a shell equilibrium.
We also notice that the CH model forms elongated droplets compared to the spherical droplets of the CHS models, and that the CH model evolves over a slower time scale, much as in the 2D case for concentration ratios near $c_1:c_2=0.5:0.5$.
However the time to equilibrium for the 3D model is reduced when compared to the 2D model at the same concentration ratio $c_1:c_2=3:7$ (7 s for CHSB $x$ in 3D, versus 47 s in 2D).
This difference may partially be explained by simple geometry.
In dimension $d$, if the volume fraction of a $c_1$ droplet is given by $\gamma = c_1/(c_1+c_2)$, then the relative radius of the inner droplet $r_1$ to the whole $r$ is given by $r_1/r = \gamma^{1/d}$.
That is to say, in three dimensions an inner $c_1$ droplet is closer to the edge of the droplet than in two dimensions.
We thus expect it to take less time to form a crescent in three dimensions.
A fairer comparison between two and three dimensions may instead be that for which the radius ratio is equal, not the volume ratio.
In this case a 2D concentration ratio of $(0.3)^{2/3}\approx 0.45$ is seen to takes less time to form a crescent (between 6 s and 35 s for $c_1:c_2=0.4:0.6$ and $0.5:0.5$ respectively).

\Cref{fig:experiment-vs-sim-big} demonstrates the qualitative correspondence between the 3D simulations and experiments of LLPS in a gelatin-PEG mixture.
We also provide videos of the experiments in the supplementary material.
The coarsening behavior, and final shapes of the particles, are in agreement.
There is a discrepancy in time scales.
The physical times suggested by the parameters in \cref{tab:dimensional-parameters} suggest much faster evolution of the simulation than the experiment. 
A time rescaling reconciles this difference and is presented in the supplementary material.

\section{Conclusions and Future Directions}
\label{sec:conclusion}

We provide several mathematical models of liquid-liquid phase separation in aqueous two-phase systems.
The simplest surface energy minimization model is able to concisely describe equilibrium droplet shapes using just three experimentally testable parameters, the surface tensions between each phase, and the relative concentration of the PEG and gelatin phases.
However this simple model does not account for the dynamics of the system during phase separation and coarsening.
To understand these dynamics, we have developed a hierarchy of three models, a ternary Cahn-Hilliard model (CH), a Cahn-Hilliard-Stokes model (CHS) to incorporate surface tension driven fluid flow, and a Cahn-Hilliard-Stokes-Boussinesq model (CHSB) to also consider buoyancy forcing.
While all models agree with the minimal energy shape and contact angles predicted by the energy minimization model, we find that they give different predictions for the same initial conditions.
For surface tension regimes corresponding to experiment ($\sigma_{1,2} \approx 0$ and $ \sigma_{1,3} \approx \sigma_{2,3}$), both the CH and CHS models equilibrate to radially symmetric shell configurations.
This contrasts with experiment, where crescents can reliably be produced.
Buoyancy forcing in the CHSB model resolves this disagreement.
It breaks the symmetry of the shell configuration, and encourages the heavier gelatin phase to sink to the bottom of the shell.
Different flow profiles for different containers can also further accelerate the phase separation process.

Several promising research directions follow.
First would be to investigate the behavior of the system for increasing separation time scales $T$.
This places more emphasis on the role of fluid dynamics but comes at steep computational cost, possibly requiring specialized time integrators \cite{LiuStabilizedSemiimplicitSpectral2015,GlasnerImprovingAccuracyConvexity2016} to allow greater time steps than the current Implicit-Explicit scheme.

The consistent failure of the CH and CHS models to achieve energy minimization is also of mathematical and physical interest.
How strongly must mixed initial conditions be perturbed to observe the emergence of minimal energy triple points?
How does the size of the necessary perturbation depend on the relative surface tensions?
Analysis inspired by  \cite{BernoffAxisymmetricSurfaceDiffusion1998} could inform how precise the manufacturing process can be,
as well as connect to findings of skin layers in models of membrane manufacture \cite{ZhouPhaseFieldSimulations2006,TreeMasstransferDrivenSpinodal2019}.
We also observe that different fluid flows strongly affect the time to crescent formation; this process may be optimized for design manufacture.

It would be scientifically valuable to explore reduced asymptotic models of phase separated low Reynolds number fluid dynamics.
The reduced equations, derived using multiple-scales matched-asymptotics as $\eps \to 0$ \cite{GlasnerDiffuseInterfaceApproach2003,MagalettiSharpinterfaceLimitCahn2013,LeeSharpInterfaceLimitsCahn2016,WangDynamicsThreePhaseTriple2017,HesterImprovingAccuracyVolume2021,HesterImprovedPhasefieldModels2020},
may provide a much more computationally efficient model of liquid-liquid phase separation, allowing comprehensive investigation of fluid dynamic effects on phase-separating polymer solutions.

Scientific advances could come from more general thermodynamic models.
Non-constant mobilities \cite{VrentasNewEquationRelating1993,ElliottCahnHilliardEquation1996} and Flory-Huggins type free energies \cite{HugginsSolutionsLongChain1941,FloryThermodynamicsHighPolymer1942} might extend the model to more widely varying temperature regimes, and further reduce free parameters of the model.
Accounting for surfactants and Marangoni stresses would also be necessary for considering droplet interactions.
Significant process on modeling cell-cell interactions has been made under a similar framework \cite{CarrilloPopulationDynamicsModel2019}.
However quantitative accuracy would necessitate much more comprehensive experimental data \cite{
ZollerStandardPressurevolumetemperatureData1995,
KoningsveldPolymerPhaseDiagrams2001,
YanagisawaPhaseSeparationBinary2014,
YamashitaDynamicsSpinodalDecomposition2018}.

We emphasize the wide applicability of our key observations. 
Microparticle manufacture, colloid engineering, and condensates in biological systems all combine fluid dynamics and phase separation.
We show that in such systems, fluid flows and forces can greatly alter the evolution and equilibria of liquid-liquid phase separation.

\matmethods{

\subsection{Experimental Determination of Densities}
\label{sec:methods}
To determine the densities of mixed homogeneous and individual phases post phase separation, we first prepared stock solutions of 4-arm PEG Acrylate (5 kDa, Advanced BioChemicals) at 20\% w/v in phosphate buffered saline (PBS) and cold water fish skin gelatin (sigma) at 10\% w/v in deionized (DI) water. We mixed these in a PEG:Gelatin ratio of 1.32:1 and added enough extra DI water until we observed a homogeneous solution. We weighed 50 $\mu$L aliquots on a mass balance in triplicate of this single phase solution. The solution was then incubated at $4$ \si{\celsius} for 15 minutes, and centrifuged to obtain separate phases. Each individual phase was weighed in 50 $\mu$L aliquots. Density was calculated by dividing the measured mass by volume of measurement and reported in g/ml units.

\subsection{Experimental Temperature-Induced Phase Separation}\cite{LeeScalableFabricationUse2022}
We prepared stock solutions of cold water fish skin gelatin (Sigma) at 10\% (w/v) in DI water and 4-arm PEG Acrylate (5 kDa, Advanced BioChemicals) at 20\% (w/v) in PBS. 
We then mixed these at the following PEG:Gelatin ratios: 3:2,1:1,1:2; and diluted each solution with DI water until we achieved a homogenous mixture. 
We then flowed this aqueous phase through a step emulsification microfluidic droplet generator at 5 $\mu$L/min as the dispersed phase.
Novec 7500 (3M) supplemented with 2\% (v/v) PicoSurf (Sphere Fluidics) at 10 $\mu$L/min was used as the continuous phase.
This produced monodisperse droplets. 
We immersed these droplets in a reservoir placed in a cold water bath at $4$ \si{\celsius}.
We imaged the droplets by acquiring a brightfield image every 30 seconds using a Photometrics Prime sCMOS camera at 10x magnification (Nikon) for 15 - 90 minutes to monitor phase separation until completion.

\subsection{Numerical methods}
\label{sec:numerics}
We implement the Cahn-Hilliard-Stokes model in the Dedalus spectral code \cite{BurnsDedalusFlexibleFramework2020}.
Dedalus automatically parses text descriptions of partial differential equations into efficient numerical solvers.
The framework is written in Python but usescompiled libraries for performance, enabling rapid prototyping and model comparisons, as well as efficient high-performance simulations.
We simulate in two and three dimensions, using Fourier projections in the horizontal directions and Chebyshev polynomials in the vertical direction, to represent no-slip, non-wetting walls.
Using basis recombination and the tau method \cite{BurnsDedalusFlexibleFramework2020} 
to enforce boundary conditions, the linear part of the system is discretized into sparse banded matrices that are parallelized over each Fourier mode using the MPI library.
A second order semi-implicit backwards-difference time stepping method iterates the linear part implicitly, and the nonlinear part explicitly.
In contrast to earlier methods involving iterative nonlinear implicit solves \cite{EyreUnconditionallyGradientStable1998,
LeeSecondorderAccurateNonlinear2008,
ChenPositivitypreservingEnergyStable2019,
ChenEnergyStableNumerical2020},
constant coefficient preconditioning alleviates equation stiffness constraints with single matrix solves (similar to 
\cite{BadalassiComputationMultiphaseSystems2003,
DongEfficientAlgorithmIncompressible2014,
DongPhysicalFormulationNumerical2015,
DongMultiphaseFlowsImmiscible2018,
ZhangDecoupledNoniterativeUnconditionally2020,
YangNewEfficientFullydecoupled2021}). 
This affords an efficient solution routine with complexity approximately linear in the degrees of freedom.
Further details are provided in the supplementary material, and the Mathematica derivation and simulation code are provided online\footnote{\href{https://github.com/ericwhester/multiphase-fluids-code}{github.com/ericwhester/multiphase-fluids-code}}.

}

\showmatmethods{} 

\acknow{This work is supported by Simons Foundation Math + X Investigator Award Number 510776. 
We acknowledge computing resources on Hoffman2 provided through the Institute for Digital Research and Education at UCLA. 
}

\showacknow{} 


\begin{thebibliography}{10}

\bibitem{XiaoReviewPreparationApplication2014}
Z Xiao, W Liu, G Zhu, R Zhou, Y Niu, A review of the preparation and
  application of flavour and essential oils microcapsules based on complex
  coacervation technology.
\newblock {\em\protect\JournalTitle{Journal of the Science of Food and
  Agriculture}} \textbf{94}, 1482--1494 (2014).

\bibitem{MaCellInspiredAllAqueousMicrofluidics2020}
Q Ma, et~al., Cell-{{Inspired All-Aqueous Microfluidics}}: {{From Intracellular
  Liquid}}\textendash{{Liquid Phase Separation}} toward {{Advanced
  Biomaterials}}.
\newblock {\em\protect\JournalTitle{Advanced Science}} \textbf{7}, 1903359
  (2020).

\bibitem{HymanLiquidLiquidPhaseSeparation2014}
AA Hyman, CA Weber, F J{\"u}licher, Liquid-{{Liquid Phase Separation}} in
  {{Biology}}.
\newblock {\em\protect\JournalTitle{Annual Review of Cell and Developmental
  Biology}} \textbf{30}, 39--58 (2014).

\bibitem{BananiBiomolecularCondensatesOrganizers2017}
SF Banani, HO Lee, AA Hyman, MK Rosen, Biomolecular condensates: Organizers of
  cellular biochemistry.
\newblock {\em\protect\JournalTitle{Nature Reviews Molecular Cell Biology}}
  \textbf{18}, 285--298 (2017).

\bibitem{BerryPhysicalPrinciplesIntracellular2018}
J Berry, CP Brangwynne, M Haataja, Physical principles of intracellular
  organization via active and passive phase transitions.
\newblock {\em\protect\JournalTitle{Reports on Progress in Physics}}
  \textbf{81}, 046601 (2018).

\bibitem{ShinLiquidPhaseCondensation2017}
Y Shin, CP Brangwynne, Liquid phase condensation in cell physiology and
  disease.
\newblock {\em\protect\JournalTitle{Science}} \textbf{357}, eaaf4382 (2017).

\bibitem{AlbertiLiquidLiquidPhase2019}
S Alberti, D Dormann, Liquid\textendash{{Liquid Phase Separation}} in
  {{Disease}}.
\newblock {\em\protect\JournalTitle{Annual Review of Genetics}} \textbf{53},
  171--194 (2019).

\bibitem{MaoPhaseBehaviorMorphology2019}
S Mao, D Kuldinow, MP Haataja, A Ko{\v s}mrlj, Phase behavior and morphology of
  multicomponent liquid mixtures.
\newblock {\em\protect\JournalTitle{Soft Matter}} \textbf{15}, 1297--1311
  (2019).

\bibitem{MaoDesigningMorphologySeparated2020}
S Mao, MS {Chakraverti-Wuerthwein}, H Gaudio, A Ko{\v s}mrlj, Designing the
  {{Morphology}} of {{Separated Phases}} in {{Multicomponent Liquid Mixtures}}.
\newblock {\em\protect\JournalTitle{Physical Review Letters}} \textbf{125},
  218003 (2020).

\bibitem{ShrinivasPhaseSeparationFluids2021}
K Shrinivas, MP Brenner, Phase separation in fluids with many interacting
  components.
\newblock {\em\protect\JournalTitle{Proceedings of the National Academy of
  Sciences}} \textbf{118} (2021).

\bibitem{ZhangPhaseFieldModelingMultiple2021}
H Zhang, Y Wu, F Wang, F Guo, B Nestler, Phase-{{Field Modeling}} of {{Multiple
  Emulsions Via Spinodal Decomposition}}.
\newblock {\em\protect\JournalTitle{Langmuir}} \textbf{37}, 5275--5281 (2021).

\bibitem{YanagisawaPhaseSeparationBinary2014}
M Yanagisawa, Y Yamashita, Sa Mukai, M Annaka, M Tokita, Phase separation in
  binary polymer solution: {{Gelatin}}/{{Poly}}(ethylene glycol) system.
\newblock {\em\protect\JournalTitle{Journal of Molecular Liquids}}
  \textbf{200}, 2--6 (2014).

\bibitem{LeeScalableFabricationUse2022}
S Lee, J {de Rutte}, R Dimatteo, D Koo, D Di~Carlo, Scalable {{Fabrication}}
  and {{Use}} of {{3D Structured Microparticles Spatially Functionalized}} with
  {{Biomolecules}}.
\newblock {\em\protect\JournalTitle{ACS Nano}} \textbf{16}, 38--49 (2022).

\bibitem{WangCountingEnzymaticallyAmplified2021}
Y Wang, et~al., Counting of enzymatically amplified affinity reactions in
  hydrogel particle-templated drops.
\newblock {\em\protect\JournalTitle{Lab on a Chip}} \textbf{21}, 3438--3448
  (2021).

\bibitem{deRutteSuspendableHydrogelNanovials2022}
J {de Rutte}, et~al., Suspendable {{Hydrogel Nanovials}} for {{Massively
  Parallel Single-Cell Functional Analysis}} and {{Sorting}}.
\newblock {\em\protect\JournalTitle{ACS Nano}} (2022).

\bibitem{ThebergeMicrodropletsMicrofluidicsEvolving2010}
AB Theberge, et~al., Microdroplets in {{Microfluidics}}: {{An Evolving
  Platform}} for {{Discoveries}} in {{Chemistry}} and {{Biology}}.
\newblock {\em\protect\JournalTitle{Angewandte Chemie International Edition}}
  \textbf{49}, 5846--5868 (2010).

\bibitem{ZarzarDynamicallyReconfigurableComplex2015}
LD Zarzar, et~al., Dynamically reconfigurable complex emulsions via tunable
  interfacial tensions.
\newblock {\em\protect\JournalTitle{Nature}} \textbf{518}, 520--524 (2015).

\bibitem{LiMicrofluidicFabricationMicroparticles2018}
W Li, et~al., Microfluidic fabrication of microparticles for biomedical
  applications.
\newblock {\em\protect\JournalTitle{Chemical Society Reviews}} \textbf{47},
  5646--5683 (2018).

\bibitem{CaiAnisotropicMicroparticlesMicrofluidics2021}
L Cai, et~al., Anisotropic {{Microparticles}} from {{Microfluidics}}.
\newblock {\em\protect\JournalTitle{Chem}} \textbf{7}, 93--136 (2021).

\bibitem{KimDropletMicrofluidicsProducing2014}
JH Kim, et~al., Droplet {{Microfluidics}} for {{Producing Functional
  Microparticles}}.
\newblock {\em\protect\JournalTitle{Langmuir}} \textbf{30}, 1473--1488 (2014).

\bibitem{ShangEmergingDropletMicrofluidics2017}
L Shang, Y Cheng, Y Zhao, Emerging {{Droplet Microfluidics}}.
\newblock {\em\protect\JournalTitle{Chemical Reviews}} \textbf{117}, 7964--8040
  (2017).

\bibitem{MaFabricationMicrogelParticles2012}
S Ma, et~al., Fabrication of {{Microgel Particles}} with {{Complex Shape}} via
  {{Selective Polymerization}} of {{Aqueous Two-Phase Systems}}.
\newblock {\em\protect\JournalTitle{Small}} \textbf{8}, 2356--2360 (2012).

\bibitem{WangHoleShellMicroparticles2013}
W Wang, et~al., Hole\textendash{{Shell Microparticles}} from {{Controllably
  Evolved Double Emulsions}}.
\newblock {\em\protect\JournalTitle{Angewandte Chemie}} \textbf{125},
  8242--8245 (2013).

\bibitem{WatanabeMicrofluidicFormationHydrogel2019}
T Watanabe, I Motohiro, T Ono, Microfluidic {{Formation}} of {{Hydrogel
  Microcapsules}} with a {{Single Aqueous Core}} by {{Spontaneous
  Cross-Linking}} in {{Aqueous Two-Phase System Droplets}}.
\newblock {\em\protect\JournalTitle{Langmuir}} \textbf{35}, 2358--2367 (2019).

\bibitem{LiuSelfOrientingHydrogelMicroBuckets2019}
Q Liu, et~al., Self-{{Orienting Hydrogel Micro-Buckets}} as {{Novel Cell
  Carriers}}.
\newblock {\em\protect\JournalTitle{Angewandte Chemie International Edition}}
  \textbf{58}, 547--551 (2019).

\bibitem{ChenMicrofluidicGeneratedBiopolymerMicroparticles2021}
Z Chen, et~al., Microfluidic-{{Generated Biopolymer Microparticles}} as {{Cargo
  Delivery Systems}}.
\newblock {\em\protect\JournalTitle{Advanced Materials Technologies}}
  \textbf{n/a}, 2100733 (2021).

\bibitem{HohenbergTheoryDynamicCritical1977}
PC Hohenberg, BI Halperin, Theory of dynamic critical phenomena.
\newblock {\em\protect\JournalTitle{Reviews of Modern Physics}} \textbf{49},
  435--479 (1977).

\bibitem{PaulsenCoalescenceBubblesDrops2014}
JD Paulsen, R Carmigniani, A Kannan, JC Burton, SR Nagel, Coalescence of
  bubbles and drops in an outer fluid.
\newblock {\em\protect\JournalTitle{Nature Communications}} \textbf{5}, 3182
  (2014).

\bibitem{ShimizuNovelCoarseningMechanism2015}
R Shimizu, H Tanaka, A novel coarsening mechanism of droplets in immiscible
  fluid mixtures.
\newblock {\em\protect\JournalTitle{Nature Communications}} \textbf{6}, 7407
  (2015).

\bibitem{SteinbacherPolymerChemistryFlow2006}
JL Steinbacher, DT McQuade, Polymer chemistry in flow: {{New}} polymers, beads,
  capsules, and fibers.
\newblock {\em\protect\JournalTitle{Journal of Polymer Science Part A: Polymer
  Chemistry}} \textbf{44}, 6505--6533 (2006).

\bibitem{ZhouPhaseFieldSimulations2006}
B Zhou, AC Powell, Phase field simulations of early stage structure formation
  during immersion precipitation of polymeric membranes in {{2D}} and {{3D}}.
\newblock {\em\protect\JournalTitle{Journal of Membrane Science}} \textbf{268},
  150--164 (2006).

\bibitem{LiPhasefieldSimulationThermally2012}
YC Li, RP Shi, CP Wang, XJ Liu, Y Wang, Phase-field simulation of thermally
  induced spinodal decomposition in polymer blends.
\newblock {\em\protect\JournalTitle{Modelling and Simulation in Materials
  Science and Engineering}} \textbf{20}, 075002 (2012).

\bibitem{TreeMasstransferDrivenSpinodal2019}
D Tree, et~al., Mass-transfer driven spinodal decomposition in a ternary
  polymer solution.
\newblock {\em\protect\JournalTitle{Soft Matter}} \textbf{15}, 4614--4628
  (2019).

\bibitem{WangPhasefieldStudyPolymerizationinduced2020}
F Wang, L Ratke, H Zhang, P Altschuh, B Nestler, A phase-field study on
  polymerization-induced phase separation occasioned by diffusion and capillary
  flow\textemdash a mechanism for the formation of porous microstructures in
  membranes.
\newblock {\em\protect\JournalTitle{Journal of Sol-Gel Science and Technology}}
  \textbf{94}, 356--374 (2020).

\bibitem{BerryRNATranscriptionModulates2015}
J Berry, SC Weber, N Vaidya, M Haataja, CP Brangwynne, {{RNA}} transcription
  modulates phase transition-driven nuclear body assembly.
\newblock {\em\protect\JournalTitle{Proceedings of the National Academy of
  Sciences}} \textbf{112}, E5237--E5245 (2015).

\bibitem{BostwickStabilityConstrainedCapillary2015}
J Bostwick, P Steen, Stability of {{Constrained Capillary Surfaces}}.
\newblock {\em\protect\JournalTitle{Annual Review of Fluid Mechanics}}
  \textbf{47}, 539--568 (2015).

\bibitem{DongMultiphaseFlowsImmiscible2018}
S Dong, Multiphase flows of {{N}} immiscible incompressible fluids: {{A}}
  reduction-consistent and thermodynamically-consistent formulation and
  associated algorithm.
\newblock {\em\protect\JournalTitle{Journal of Computational Physics}}
  \textbf{361}, 1--49 (2018).

\bibitem{AndersonDiffuseInterfaceMethodsFluid1998}
DM Anderson, GB McFadden, AA Wheeler, Diffuse-{{Interface Methods}} in {{Fluid
  Mechanics}}.
\newblock {\em\protect\JournalTitle{Annual Review of Fluid Mechanics}}
  \textbf{30}, 139--165 (1998).

\bibitem{JacqminCalculationTwoPhaseNavier1999}
D Jacqmin, Calculation of {{Two-Phase Navier}}\textendash{{Stokes Flows Using
  Phase-Field Modeling}}.
\newblock {\em\protect\JournalTitle{Journal of Computational Physics}}
  \textbf{155}, 96--127 (1999).

\bibitem{KimPhaseFieldModeling2005}
J Kim, J Lowengrub, Phase field modeling and simulation of three-phase flows.
\newblock {\em\protect\JournalTitle{Interfaces and Free Boundaries}}
  \textbf{7}, 435--466 (2005).

\bibitem{BoyerStudyThreeComponent2006}
F Boyer, C Lapuerta, Study of a three component {{Cahn-Hilliard}} flow model.
\newblock {\em\protect\JournalTitle{ESAIM: Mathematical Modelling and Numerical
  Analysis}} \textbf{40}, 653--687 (2006).

\bibitem{BoyerHierarchyConsistentNcomponent2014}
F Boyer, S Minjeaud, Hierarchy of consistent n-component
  {{Cahn}}\textendash{{Hilliard}} systems.
\newblock {\em\protect\JournalTitle{Mathematical Models and Methods in Applied
  Sciences}} \textbf{24}, 2885--2928 (2014).

\bibitem{AlikakosConvergenceCahnHilliardEquation1994}
ND Alikakos, PW Bates, X Chen, Convergence of the {{Cahn-Hilliard}} equation to
  the {{Hele-Shaw}} model.
\newblock {\em\protect\JournalTitle{Archive for Rational Mechanics and
  Analysis}} \textbf{128}, 165--205 (1994).

\bibitem{CahnFreeEnergyNonuniform1958}
JW Cahn, JE Hilliard, Free {{Energy}} of a {{Nonuniform System}}. {{I}}.
  {{Interfacial Free Energy}}.
\newblock {\em\protect\JournalTitle{The Journal of Chemical Physics}}
  \textbf{28}, 258--267 (1958).

\bibitem{TjahjadiSatelliteSubsatelliteFormation1992}
M Tjahjadi, HA Stone, JM Ottino, Satellite and subsatellite formation in
  capillary breakup.
\newblock {\em\protect\JournalTitle{Journal of Fluid Mechanics}} \textbf{243},
  297 (1992).

\bibitem{StoneDynamicsDropDeformation1994}
HA Stone, Dynamics of {{Drop Deformation}} and {{Breakup}} in {{Viscous
  Fluids}}.
\newblock {\em\protect\JournalTitle{Annual Review of Fluid Mechanics}}
  \textbf{26}, 65--102 (1994).

\bibitem{LiuStabilizedSemiimplicitSpectral2015}
F Liu, J Shen, Stabilized semi-implicit spectral deferred correction methods
  for {{Allen}}\textendash{{Cahn}} and {{Cahn}}\textendash{{Hilliard}}
  equations.
\newblock {\em\protect\JournalTitle{Mathematical Methods in the Applied
  Sciences}} \textbf{38}, 4564--4575 (2015).

\bibitem{GlasnerImprovingAccuracyConvexity2016}
K Glasner, S Orizaga, Improving the accuracy of convexity splitting methods for
  gradient flow equations.
\newblock {\em\protect\JournalTitle{Journal of Computational Physics}}
  \textbf{315}, 52--64 (2016).

\bibitem{BernoffAxisymmetricSurfaceDiffusion1998}
AJ Bernoff, AL Bertozzi, TP Witelski, Axisymmetric {{Surface Diffusion}}:
  {{Dynamics}} and {{Stability}} of {{Self-Similar Pinchoff}}.
\newblock {\em\protect\JournalTitle{Journal of Statistical Physics}}
  \textbf{93}, 725--776 (1998).

\bibitem{GlasnerDiffuseInterfaceApproach2003}
K Glasner, A diffuse interface approach to {{Hele Shaw}} flow.
\newblock {\em\protect\JournalTitle{Nonlinearity}} \textbf{16}, 49--66 (2003).

\bibitem{MagalettiSharpinterfaceLimitCahn2013}
F Magaletti, F Picano, M Chinappi, L Marino, CM Casciola, The sharp-interface
  limit of the {{Cahn}}\textendash{{Hilliard}}/{{Navier}}\textendash{{Stokes}}
  model for binary fluids.
\newblock {\em\protect\JournalTitle{Journal of Fluid Mechanics}} \textbf{714},
  95--126 (2013).

\bibitem{LeeSharpInterfaceLimitsCahn2016}
AA Lee, A M{\"u}nch, E S{\"u}li, Sharp-{{Interface Limits}} of the
  {{Cahn--Hilliard Equation}} with {{Degenerate Mobility}}.
\newblock {\em\protect\JournalTitle{SIAM Journal on Applied Mathematics}}
  \textbf{76}, 433--456 (2016).

\bibitem{WangDynamicsThreePhaseTriple2017}
D Wang, XP Wang, YG Wang, The {{Dynamics}} of {{Three-Phase Triple Junction}}
  and {{Contact Points}}.
\newblock {\em\protect\JournalTitle{SIAM Journal on Applied Mathematics}}
  \textbf{77}, 1805--1826 (2017).

\bibitem{HesterImprovingAccuracyVolume2021}
EW Hester, GM Vasil, KJ Burns, Improving accuracy of volume penalised
  fluid-solid interactions.
\newblock {\em\protect\JournalTitle{Journal of Computational Physics}}
  \textbf{430}, 110043 (2021).

\bibitem{HesterImprovedPhasefieldModels2020}
EW Hester, LA Couston, B Favier, KJ Burns, GM Vasil, Improved phase-field
  models of melting and dissolution in multi-component flows.
\newblock {\em\protect\JournalTitle{Proceedings of the Royal Society A:
  Mathematical, Physical and Engineering Sciences}} \textbf{476}, 20200508
  (2020).

\bibitem{VrentasNewEquationRelating1993}
JS Vrentas, CM Vrentas, A new equation relating self-diffusion and mutual
  diffusion coefficients in polymer-solvent systems.
\newblock {\em\protect\JournalTitle{Macromolecules}} \textbf{26}, 6129--6131
  (1993).

\bibitem{ElliottCahnHilliardEquation1996}
CM Elliott, H Garcke, On the {{Cahn}}\textendash{{Hilliard Equation}} with
  {{Degenerate Mobility}}.
\newblock {\em\protect\JournalTitle{SIAM Journal on Mathematical Analysis}}
  \textbf{27}, 404--423 (1996).

\bibitem{HugginsSolutionsLongChain1941}
ML Huggins, Solutions of {{Long Chain Compounds}}.
\newblock {\em\protect\JournalTitle{The Journal of Chemical Physics}}
  \textbf{9}, 440--440 (1941).

\bibitem{FloryThermodynamicsHighPolymer1942}
PJ Flory, Thermodynamics of {{High Polymer Solutions}}.
\newblock {\em\protect\JournalTitle{The Journal of Chemical Physics}}
  \textbf{10}, 51--61 (1942).

\bibitem{CarrilloPopulationDynamicsModel2019}
JA Carrillo, H Murakawa, M Sato, H Togashi, O Trush, A population dynamics
  model of cell-cell adhesion incorporating population pressure and density
  saturation.
\newblock {\em\protect\JournalTitle{Journal of Theoretical Biology}}
  \textbf{474}, 14--24 (2019).

\bibitem{ZollerStandardPressurevolumetemperatureData1995}
P Zoller, {\em Standard Pressure-Volume-Temperature Data for Polymers}.
\newblock ({Technomic Pub. Co.}, {Lancaster, PA}), (1995).

\bibitem{KoningsveldPolymerPhaseDiagrams2001}
R Koningsveld, {\em Polymer Phase Diagrams / {{Ronald Koningsveld}}, {{Walter
  H}}. {{Stockmayer}} and {{Erik Nies}}.}
\newblock ({University Press}, {Oxford}), (2001).

\bibitem{YamashitaDynamicsSpinodalDecomposition2018}
Y Yamashita, M Yanagisawa, M Tokita, Dynamics of {{Spinodal Decomposition}} in
  a {{Ternary Gelling System}}.
\newblock {\em\protect\JournalTitle{Gels}} \textbf{4}, 26 (2018).

\bibitem{BurnsDedalusFlexibleFramework2020}
KJ Burns, GM Vasil, JS Oishi, D Lecoanet, BP Brown, Dedalus: {{A}} flexible
  framework for numerical simulations with spectral methods.
\newblock {\em\protect\JournalTitle{Physical Review Research}} \textbf{2},
  023068 (2020).

\bibitem{EyreUnconditionallyGradientStable1998}
DJ Eyre, Unconditionally {{Gradient Stable Time Marching}} the {{Cahn-Hilliard
  Equation}}.
\newblock {\em\protect\JournalTitle{MRS Online Proceedings Library (OPL)}}
  \textbf{529} (1998/ed).

\bibitem{LeeSecondorderAccurateNonlinear2008}
HG Lee, J Kim, A second-order accurate non-linear difference scheme for the
  {{N}} -component {{Cahn}}\textendash{{Hilliard}} system.
\newblock {\em\protect\JournalTitle{Physica A: Statistical Mechanics and its
  Applications}} \textbf{387}, 4787--4799 (2008).

\bibitem{ChenPositivitypreservingEnergyStable2019}
W Chen, C Wang, X Wang, SM Wise, Positivity-preserving, energy stable numerical
  schemes for the {{Cahn-Hilliard}} equation with logarithmic potential.
\newblock {\em\protect\JournalTitle{Journal of Computational Physics: X}}
  \textbf{3}, 100031 (2019).

\bibitem{ChenEnergyStableNumerical2020}
W Chen, C Wang, S Wang, X Wang, SM Wise, Energy {{Stable Numerical Schemes}}
  for {{Ternary Cahn-Hilliard System}}.
\newblock {\em\protect\JournalTitle{Journal of Scientific Computing}}
  \textbf{84}, 27 (2020).

\bibitem{BadalassiComputationMultiphaseSystems2003}
V Badalassi, H Ceniceros, S Banerjee, Computation of multiphase systems with
  phase field models.
\newblock {\em\protect\JournalTitle{Journal of Computational Physics}}
  \textbf{190}, 371--397 (2003).

\bibitem{DongEfficientAlgorithmIncompressible2014}
S Dong, An efficient algorithm for incompressible {{N-phase}} flows.
\newblock {\em\protect\JournalTitle{Journal of Computational Physics}}
  \textbf{276}, 691--728 (2014).

\bibitem{DongPhysicalFormulationNumerical2015}
S Dong, Physical formulation and numerical algorithm for simulating {{N}}
  immiscible incompressible fluids involving general order parameters.
\newblock {\em\protect\JournalTitle{Journal of Computational Physics}}
  \textbf{283}, 98--128 (2015).

\bibitem{ZhangDecoupledNoniterativeUnconditionally2020}
J Zhang, X Yang, Decoupled, non-iterative, and unconditionally energy stable
  large time stepping method for the three-phase {{Cahn-Hilliard}} phase-field
  model.
\newblock {\em\protect\JournalTitle{Journal of Computational Physics}}
  \textbf{404}, 109115 (2020).

\bibitem{YangNewEfficientFullydecoupled2021}
X Yang, A new efficient fully-decoupled and second-order time-accurate scheme
  for {{Cahn}}\textendash{{Hilliard}} phase-field model of three-phase
  incompressible flow.
\newblock {\em\protect\JournalTitle{Computer Methods in Applied Mechanics and
  Engineering}} \textbf{376}, 113589 (2021).

\end{thebibliography}

\end{document}



\maketitle

\SItext

\section{Minimal Surface Energy Solutions}
\label{sec:equilibrium}
At equilibrium the system should satisfy minimization of the surface energy. 
This amounts to minimizing surface areas, weighted according to the surface tension $\sigma_{i,j}$ of the respective $i,j$ interface.
This condition is satisfied by surfaces that are part of a sphere \cite{BostwickStabilityConstrainedCapillary2015},
From which we can derive explicit forms for the minimal energy shape as a function of the surface tensions $\sigma_{i,j}$, and the relative volumes of the $c_1$ and $c_2$ phases.
The contact angle condition requires
	\begin{align}
	\frac{\sigma_{1,2}}{\sin \theta_3} = 
	\frac{\sigma_{1,3}}{\sin \theta_2} &= 
	\frac{\sigma_{2,3}}{\sin \theta_1}, &
	\theta_1 + \theta_2 + \theta_3 &= 2 \pi.
	\end{align}
If any surface tension dominates the sum of the remaining two, then this system is no longer solvable, and no stable triple point may form.
We now provide explicit expression for the shape of the solution as a function of surface energies and concentration ratio, for a total volume $\int_\Omega c_1 + c_2 dV = 1$.

\subsection{Separated spheres} ($\chi_3 < 0$):
If the surface tensions of the 1,2 interface is too large, then the two phases coalesce into two droplets separated by the third phase.
For a total volume equal to 1, and volume of the first phase given by $v_1$, then the radii $r_1,r_2$ of the droplets, and the total surface energy $e$ are given by
	\begin{align}
	r_1 &= \br{\frac{3}{4\pi} v_1}^{1/3}, &
	r_2 &= \br{\frac{3}{4\pi} (1-v_1)}^{1/3}, &
	e &= \br{6^2\pi v_1^2}^{1/3} \sigma_{1,3} + \br{6^2\pi (1-v_1)^2}^{1/3} \sigma_{2,3}.
	\end{align}

\subsection{1-in-2 shell} ($\chi_2 < 0$): 
If the 1,3 surface tension is too large, then the system chooses a 1 sphere inside a 2 shell, with respective radii $r_1,r_2$ and total surface energy given by
	\begin{align}
	r_1 &= \br{\frac{3}{4\pi} v_1}^{1/3}, &
	r_2 &= \br{\frac{3}{4\pi}}^{1/3}, &
	e &= \br{6^2\pi v_1^2}^{1/3} \sigma_{1,2} + \br{6^2\pi}^{1/3} \sigma_{2,3}.
	\end{align}

\subsection{2-in-1 shell} ($\chi_1 < 0$):
If instead the 2,3 surface tension is too large, then the system chooses a 2 sphere inside a 1 shell, with respective radii $r_1,r_2$ and total surface energy given by
	\begin{align}
	r_1 &= \br{\frac{3}{4\pi} (1-v_1)}^{1/3}, &
	r_2 &= \br{\frac{3}{4\pi}}^{1/3}, &
	e &= \br{6^2\pi (1-v_1)^2}^{1/3} \sigma_{1,2} + \br{6^2\pi}^{1/3} \sigma_{1,3}.
	\end{align}

\subsection{Crescent particle} 
\begin{figure}[H]
\centering
    \includegraphics[width=.2\linewidth]{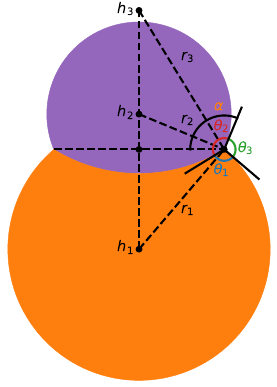}
	\label{fig:crescent-diagram}
    \caption{Schematic of crescent particle calculations.}
\end{figure}

($\chi_3 < 0$):
The most complex case occurs when all wetting parameters are positive, and a crescent particle forms the minimum energy shape.
Here, we define several auxiliary quantities (shown in \cref{fig:crescent-diagram}) to analyze the energy calculation.
The shape is given by three spherical caps for the 2-3 (top, cap 1), 1-3 (bottom, cap 2), and 1-2 (middle, cap 3) phases.
Each spherical cap intersects the circle with radius $\ell$, and is described by its radius $r_i$, and the angle it subtends with the $z=0$ plane.
One can define each of these angles in terms of $\alpha$, and two of the remaining contact angles, $\theta_1,\theta_2$.
Changing the angle $\alpha$ while holding $\theta_i$ constant alters the relative volumes of each shape.
The total surface area is then the sum of the areas of each spherical cap, and the total volume is the sum of the top and bottom spherical cap.
A spherical cap with angle $\phi$ and width $2\ell$ has a radius $r$, a height to the center $h$, a surface area $s$, and a volume $v$ given by
	\begin{align}
	h &= \frac{\ell}{\tan \phi},&
	r &= \frac{\ell}{\sin \phi},&
	s &= 2 \pi r^2 (1 - \cos \phi),&
	v &= \frac{\pi}{3} r^3 (2 + \cos \phi)(1-\cos \phi)^2.
	\end{align}
The angle for each spherical cap $\phi_i$ is given in terms of $\alpha,\theta_1,\theta_2,\theta_3$ by
	\begin{align}
	\phi_1 &= \alpha,&
	\phi_2 &= \alpha-\theta_1 - \theta_2,&
	\phi_3 &= \alpha-\theta_1.
	\end{align}
The volume of each phase $v_1,v_2$ therefore depends on the width $\ell$ and the angle $\alpha$.
This relationship must then be numerically inverted to determine $\ell(v_1,v_2), \alpha(v_1,v_2)$.
From this, and the relationship for the surface area of each cap, the surface energy of a crescent with defined contact angles $\theta_i$ and volumes $v_1,v_2$ can then be calculated.

\section{Model Derivation}

\subsection{Ternary Cahn-Hilliard: Dynamic surface energy minimization}
\label{sec:cahn-hilliard}
The surface energy minimization model only predicts equilibrium shapes.
But droplet formation involves thermally induced phase separation, an inherently nonequilibrium process.
To model dynamic phase separation from an initially mixed model we use a ternary generalization of the Cahn-Hilliard \cite{CahnFreeEnergyNonuniform1958} (or model B \cite{HohenbergTheoryDynamicCritical1977}) equation.
Multiphase Cahn-Hilliard models are derived from an energy functional
	\begin{align*}
	F &= \int_\Omega f(c) + \sum_{ij} l_{ij}\nabla c_i \cdot \nabla c_j \, dV,
	\end{align*}
where the gradient energy tensor $l_{ij}$ is positive definite.
The phases $c_i$ evolve according to conservation laws with fluxes given by gradients in the chemical potential $\mu_i$
	\begin{align*}
	\pd_t c_i &= \nabla \cdot \br{\sum_j m_{ij} \nabla \mu_j},
	\end{align*}
where $m_{ij}$ are the components of the mobility tensor and $\mu_i$ are the variational derivative of the free energy $F$ with respect to concentration $c_i$
	\begin{align*}
	\mu_i &= \frac{\delta F}{\delta c_i} = \frac{\pd}{\pd c_i} f(c) - \sum_j l_{ij} \nabla^2 c_j.
	\end{align*}
In general the mobility tensor $m$ and gradient energy tensor $l$ depend on concentrations $c_i$, but here we assume they are constant. 
It is straightforward to show that the system will {conserve} mass ($\pd_t \int_\Omega c_i \, dV = 0$), {dissipate} energy ($\pd_t F \leq 0 $), and {balance} concentrations ($\sum_i \pd_t c_i = 0$), provided the system has zero-flux boundary conditions, the mobility tensor is
positive semi-definite, and the 
nullspace constraint $\sum_i m_{ij} = 0$.

To ensure quantitative agreement with experiment, we must tune the surface energy of each interface at equilibrium.
Here, we adopt the form given in \cite{DongMultiphaseFlowsImmiscible2018}, wherein
\begin{align*}
	F &= \beta \br{\sum_i \chi_i c_i^2(1-c_i)^2 - \frac{\bar{\eps}^2}{2} \sum_{i,j} \sigma_{i,j} \nabla c_i \cdot \nabla c_j},\\
	m_{ij} &= \begin{cases}
		2m_0 & i = j,\\
		-m_0 & i \neq j,
	\end{cases} \qquad
	l_{ij} = - \frac{1}{2}\beta \bar{\eps}^2 \sigma_{i,j},
\end{align*}
where we define the energy scale $\beta = {3}/{\sqrt{2} \bar{\eps}}$ in terms of the equilibrium interface thickness $\bar{\eps}$, and the wetting parameters $\chi_i$.
It is possible to use asymptotic arguments to show that the equilibrium ternary Cahn-Hilliard model reduces to the surface-energy minimization model in the limit $\bar{\eps} \to 0$ \cite{AlikakosConvergenceCahnHilliardEquation1994}.

\subsection{Incorporating surface tension and buoyancy for small scale fluids}
\label{sec:CHSB}
While the ternary Cahn-Hilliard model captures the dynamics of surface-energy minimization, it neglects the critical role of fluid flow on phase separation.
We incorporate this following the work of \cite{DongMultiphaseFlowsImmiscible2018} (which generalizes \cite{AndersonDiffuseInterfaceMethodsFluid1998,JacqminCalculationTwoPhaseNavier1999,KimPhaseFieldModeling2005,BoyerStudyThreeComponent2006,BoyerHierarchyConsistentNcomponent2014}).
The Cahn-Hilliard equations have an added advection term from a local averaged fluid velocity $u$,
	\begin{align*}
	\pd_t c_i +  u\cdot \nabla c - \sum_{j} \nabla \cdot \br{m_{ij} \nabla \mu_j} &= 0.
	\end{align*}
The velocity evolves according to the balance of inertia, pressure, viscosity, momentum flux, surface tension, and buoyancy
    \begin{align*}
    \rho(\pd_t u + u\cdot \nabla u) + \nabla {p} - \nabla \cdot \br{\eta (\nabla u+ \nabla u^\top)} + J\cdot \nabla u \\
    	 \quad - \beta \sum_{i} \mu_i \nabla c_i + (\rho-\rho_0) g\hat{g} &= 0,
    \end{align*}
where the surface tension is redefined in terms of $\mu_i$, the gravitational acceleration is given by $g \hat{g}$, and the density $\rho$, dynamic viscosity $\eta$, and flux $J$ are given by
    \begin{align*}
    \rho(c) &= \rho_0 + \sum_i d\rho_i c_i, &	
	\eta(c) &= \sum_i \eta_i c_i, &
	J &= -\sum_{i,j} \rho_i m_{ij} \nabla \mu_j.
	\end{align*}
The flux $J$ is designed to ensure energy conservation \cite{DongMultiphaseFlowsImmiscible2018}.
Fluid incompressibility requires
	\begin{align*}
	\nabla \cdot u = 0.
	\end{align*}

\subsection{Dimensionless numbers}
We redefine dimensional quantities in terms of their dimensional scale and dimensionless variables to isolate the relevant control parameters
	\begin{align*}
	x &= L x', &
	t &= T t', &
	\bar{\eps} &= L \eps, \\
	u &= U u', &
	p &= P p', & 
	\rho_i &= \rho_0 + \Delta\rho d\rho_i', \\
	\eta_i &= \rho_0 \nu_0 \nu_i', &
	\sigma_{i,j} &= \sigma_0 \sigma_{i,j}', &
	m_{ij} &= m_0 m_{ij}'.
	\end{align*}
Here $\nu_0$ represents the average kinematic viscosity of the solution.
We choose the interfacial energy scale $\sigma_0$ so that the dimensionless surface tensions sum to one,
	\begin{align*}
	\sigma_0 &= \sigma_{1,2} + \sigma_{1,3} + \sigma_{2,3}.
	\end{align*}
We then choose the following scalings
	\begin{align*}
	U &= \frac{L}{T}, &
	P &= \We^{-1}\rho_0 U^2, &
	g &= \Bo\br{\frac{\sigma_0}{\Delta \rho L^2}},\\
	\nu_0 &= \Ca \br{\frac{\sigma_0}{\rho_0 U}} , &
	T &= \br{\frac{\rho_0 L^3}{\sigma_0 \We}}^{1/2},&
	m_0 &= \frac{\sqrt{2}}{3} \eps \frac{L^2}{T\sigma_0}.
	\end{align*}
These lead to the dimensionless interface thickness $\eps$, the capillary number $\Ca$, Bond number $\Bo$ and Weber number $\We$, which compare viscosity, buoyancy, and inertia with surface tension forces, respectively:
	\begin{align*}
	\eps &= \frac{\bar{\eps}}{L}, &
	\Ca &= \frac{\rho_0 \nu_0 L}{\sigma_0 T}, &
	\Bo &= \frac{\Delta \rho \, g L^2}{\sigma_0}, &
	\We &= \frac{\rho_0 L^3}{\sigma_0 T^2}.
	\end{align*}
Droplet formation dynamics depends on the relative sizes of these numbers.

\subsection{The Cahn-Hilliard-Stokes-Boussinesq model}
Dropping primes for dimensionless quantities, neglecting inertia, and applying the Boussinesq approximation (neglecting density variation aside from the buoyancy term) with a constant viscosity leads to the following system of equations
	\begin{align*}
	\pd_t c_i + u \cdot \nabla c_i - \sum_{j}m_{ij}\Delta \mu_j &= 0,\\
	\nabla p - \Ca \nabla^2 u - \Bo \rho \, \hat{g} &= \frac{3}{\sqrt{2}}\frac{1}{\eps} \sum_{i}\mu_i \nabla c_i,\\
	\nabla \cdot u &= 0,
	\end{align*}
where now
	\begin{align*}
	\mu_i &= 2\chi_i c_i(1-c_i)(1-2c_i) + \eps^2 \sum_j \sigma_{i,j}\Delta c_j,\\
	m_{ij}' &= \begin{cases}
		2 & i = j,\\
		-1 & i \neq j.
	\end{cases}
	\end{align*}
The derivation and nondimensionalization of the equations is provided in a Mathematica notebook available at \\\href{https://github.com/ericwhester/multiphase-fluids-code}{github.com/ericwhester/multiphase-fluids-code}.

\section{Model Discretization}

\subsection{Spatial domain and boundary conditions}
These PDEs are solved in either a 2D $(x,z) \in [-1,1]\times[-1,1]$ or 3D $(x,y,z) \in [-1,1]\times[-1,1]\times[-1,1]$ spatial domain.
Periodic boundary conditions are applied in the horizontal directions
    \begin{align}
    c_i|_{x=-1} &= c_i|_{x=1}, &
    c_i|_{y=-1} &= c_i|_{y=1}, &
    u|_{x=-1} &= u|_{x=1}, &
    u|_{y=-1} &= u|_{y=1}.
    \end{align}
At the top and bottom boundary, we specify no-flux, non-wetting, no-slip boundary conditions
    \begin{align}
    u|_{z=1} = \partial_z \mu_i|_{z=1} = c_i|_{z=1} = u|_{z=-1} = \partial_z \mu_i|_{z=-1} = c_i|_{z=-1} = 0,
    \end{align}
where we again only solve and apply conditions for phases $i=1,2$.

\subsection{Dedalus reformulation}
All Dedalus initial value problems are ultimately put into the following form
    \begin{align}
    M. \pd_t X + L. X &= F(X),
    \end{align}
where $X$ is a vector of state variables, $M$ and $L$ are linear operators, and $F$ may be a possibly nonlinear operator applied to $X$. 
We recapitulate the key steps used to discretize PDEs in the Dedalus framework (see \cite{BurnsDedalusFlexibleFramework2020} for further details).
\subsection{Time stepping}
Temporal discretization for multistep methods is done by evaluating a linear combination of each terms at different time steps
    \begin{align}
    M.\pd_t X &\approx \sum_{i=0}^m M.(a_i X^{n-i}), &
    L X &\approx \sum_{i=0}^m L.(b_i X^{n-i}), &
    F(X) &\approx \sum_{i=1}^m c_i F(X^{n-i}),
    \end{align}
where $X^{n-i}$ is the state vector evaluated at time step $n-i$, and $a,b,$ and $c$ are vectors for linear combinations of the time derivative term $M. \pd_t X$, the linear term $L.X$, and the nonlinear term $F(X)$.
This is converted into a matrix solve for $X^n$
    \begin{align}
    \br{a_0 M + b_0 L}. X^n &= \sum_{i=1}^m \br{c_i F(X^{n-i}) - a_i M.X^{n-i} - b_i L. X^{n-i}}.
    \end{align}
In our case, we use a second order backwards difference formula where
    \begin{align}
    a &= \left[\frac{1+2 w_1}{(1+w_1)k_1}, -\frac{1+w_1}{k_1}, \frac{w_1^2}{(1+w_1)k_1}\right],&
    b &= \left[1, 0, 0\right],&
    c &= \left[0, 1+w_1, -w_1\right],
    \end{align}
where $k_1$ is the most recent timestep size, and $k_0$ is the timestep size before that, and $w_1 = k_1/k_0$.

\subsection{Spatial discretization}
The state vector consists of a spectral discretization of the vector of system variables.
In our 3D simulations we discretize using rescaled Fourier series $e^{i\pi jx}, e^{i\pi ky}$ in the horizontal $x,y$ dimensions, and Chebyshev polynomials $T_\ell(2 z - 1)$ in the vertical direction.
The $m$th variable in the state vector can thus be written 
    \begin{align}
    X_m &= \sum_{j=0}^{N_x} \sum_{k=0}^{N_y}\sum_{\ell=0}^{N_z} X_{m, j, k, \ell} e^{i \pi j x} e^{ i \pi k y} T_\ell(2 z - 1) + \text{c.c.}
    \end{align}
where $\text{c.c.}$ refers to the complex conjugate. Note that we do not add the complex conjugate of real terms in this sum ($j=0$ and/or $k=0$).
The components of each matrix $M,L$ are given by their action on the corresponding basis function $e^{i \pi j x} e^{ i \pi k y} T_\ell(2 z - 1)$ when projected against the new basis functions $e^{i \pi j x} e^{ i \pi k y} U_\ell(2 z - 1)$, where $U_\ell$ is the $\ell$th Chebyshev polynomial of the second kind.
For linear operators consisting of multiplication by a constant, or $x,y,z$ derivatives, this operation is sparse.
Indeed, these operators are diagonalized by Fourier modes, and so are block separable for the Fourier modes.
This means the matrix solves can be parallelized over each Fourier mode, where we solve a matrix for the Chebyshev coefficients of all the variables (ordered first by Chebyshev mode and second by variable, giving a matrix bandwidth proportional to the number of variables).
This requires adding new variables to the system in order to ensure no left hand side contains more than one $z$ derivative; for more details see \cite{BurnsDedalusFlexibleFramework2020}.

\subsection{First order formulation}
In our case, the state vector $X$ consists of the variables
    \begin{align}
    X &= [c_1, c_2, c_{1,z}, c_{2,z}, \nabla^2 c_1, \nabla^2c_2, m_1, m_2, m_{1,z}, m_{2,z}, u_x, u_y, u_z, u_{x,z}, u_{y,z}, u_{z,z}, p].
    \end{align}
The mathematical system can be written as
    \begin{align}
    \pd_t c_i - (\pd_x^2 + \pd_y^2) m_i + \pd_z m_{i,z}) &= - \br{u_x \pd_x c_i + u_y \pd_y c_i + u_z c_{i,z} }, & i &= \{1,2\},\\
    \nabla^2 c_i - \br{(\pd_x^2 + \pd_y^2) c_i + \pd_z c_{i,z}} &= 0, & i &= \{1,2\},\\
    m_i - m_{i,L} &= m_{i,R}, & i &= \{1,2\},\\
    \pd_K p - \Ca ((\pd_x^2 + \pd_y^2)u_K + \pd_z u_{K,z}) &= \text{surf}_K + \text{buoyancy}_K,&  K&=\{x,y,z\},\\
    \pd_x u_x + \pd_y u_y + u_{z,z} &= 0,\\
    \pd_z f - f_z &= 0, & f &= \{c_1,c_2,m_1,m_2,u_x,u_y,u_z\},
    \end{align}
where we define
    \begin{align}
    m_{1,L} &= ((4 \chi_1 + 2 \chi_3)c_1 + (2\chi_3 - 2 \chi_2) c_2 + \eps^2\br{(-\sigma_{1,2} - 3\sigma_{1,3}+\sigma_{2,3}) \nabla^2 c_1 + 2(\sigma_{1,2}-\sigma_{1,3})\nabla^2 c_2},\\
    m_{2,L} &= ((2 \chi_3 - 2 \chi_1)c_1 + (4\chi_2 + 2 \chi_3) c_2 + \eps^2\br{2(\sigma_{1,2} - \sigma_{2,3}) \nabla^2 c_1 + (-\sigma_{1,2} + \sigma_{1,3} - 3 \sigma_{2,3})\nabla^2 c_2},\\
    m_{1,R} &= 2\chi_1 q(c_1) - \chi_2 q(c_2) - \chi_3 q(c_3),\\
    m_{2,R} &= -\chi_1 q(c_1) +2 \chi_2 q(c_2) - \chi_3 q(c_3),\\
    q(c) &= -6c^2 + 4c^3,\\
    c_3 &= 1 - c_1 - c_2,\\
    \text{surf}_K &= \sum_{i=1}^3\mu_i \pd_K c_i, \quad K=\{x,y,z\},\\
    \mu_i &= \chi_i g'(c_i) + \eps^2 \sum_{j=1}^3 \sigma_{i,j} \nabla^2 c_j, \quad i=\{1,2,3\},\\
    g'(c) &= 2 c(1-c)(1-2c),\\
    \text{buoyancy}_K &= \begin{cases}
        [0,0,0] & \text{CHS},\\
        [0,0,-\Bo] & \text{CHSB } z,\\
        [-\Bo,0,0] & \text{CHSB } x
    \end{cases}
    \end{align}
For 2D simulations we remove all $y$ dependent terms, and for the CH model we do not solve for the fluid velocity.

\subsection{Boundary conditions}
The matrices described by the discretization of the governing equations are not of full rank, and so cannot be solved.
This is because we require vertical boundary conditions to determine unique solutions to the problem.
Our boundary conditions at $z=\pm 1$ amount to homogeneous Dirichlet conditions on $c_1, c_2, \mu_1, \mu_2, u_x, u_y, u_z$.
These conditions can be expressed by a linear combination of the Chebyshev coefficients of the state vector for each Fourier mode.
To enforce these conditions, for each Fourier mode we replace the matrix entries for the final Chebyshev mode of each state variable with a boundary condition row.
This technique thus solves an approximation to the original differential equation.
We can view our problem as solving an approximate PDE that is perturbed by an additional Chebyshev polynomial of the second kind (of order $N_z$), $U_{N_z}$ for each variable and Fourier mode.
This exact solution to an approximate problem is known as the Tau method \cite{BurnsDedalusFlexibleFramework2020}.

\subsection{Gauge constraints}
There is a remaining degeneracy due to the gauge constraint on the pressure.
The constant mode for the vertical velocity is overdetermined by top and bottom no-slip boundary conditions when combined with the divergence-free constraint.
The corresponding constant mode for the pressure is also unconstrained by the equations.
We render the matrix of full rank by replacing the no-slip boundary condition for the vertical velocity at $z=-1$ with a gauge constraint $p = 0$ for the constant pressure mode.

\subsection{Basis recombination}
Derivatives of Chebyshev polynomials $T_n$ are not sparse when expressed in terms of Chebyshev polynomials.
If we instead express derivatives in terms of Chebyshev polynomials of the second kind $U_k$, then differentiation, as well as the identity, become sparse operations
    \begin{align}
    T_n(z) &= \frac{1}{2}(U_n - U_{n-2}), &
    T'_n(z) &= n U_{n-1}(z).
    \end{align}
This amounts to left preconditioning the system.
Boundary conditions pose a second problem, as they contribute dense rows to the system matrix.
If however, we express our solution in terms of the recombined `Dirichlet' basis
    \begin{align}
    \label{eq:dirichlet}
    D_n(z) &= T_n(z) - T_{n-1}(z),
    \end{align}
then Dirichlet boundary conditions only involve the first two modes.
This amounts to right preconditioning of the system.

Thus, the full solution procedure involves first, for each Fourier mode, building the matrix over all variables and Chebyshev modes, as expressed going from the $D_n$ basis to the $U_n$ basis. 
This matrix is sparse and banded, and an LU factorization is performed for fast solves.
This is done once at the start of the simulation if time steps are not varied.
Then, for each time step
\begin{enumerate}
    \item Calculate the right hand side in terms of the $U_n$ basis. Perform Fast Fourier Transforms (and Discrete Cosine Transforms) to evaluate suboperations in the sparsest basis (multiplication is performed on grid values, differentiation is performed on spectral coefficients, addition and scalar multiplication can be performed in either basis).
    \item Solve for the $D_n$ coefficients using the sparse LU factorization of the system matrix over each Fourier mode.
    \item Solve for the $T_n$ coefficients using a sparse solve according to \cref{eq:dirichlet}.
\end{enumerate}
Each time step involves some number of Fast Fourier Transforms, and the application or solution of sparse banded matrices, all of which are either linear or log-linear complexity in the number of degrees of freedom.

\section{Simulation Parameters}
Estimated dimensional parameters for our system are given in \cref{tab:dimensional-parameters}.
The corresponding nondimensional parameters are provided in \cref{tab:nondimensional-parameters}.
The numerical parameters used in the discretizations are given in \cref{tab:numerical-parameters}.

\begin{table}[hbt]
	\centering
	\begin{tabular}{lcl}
    Name & Symbol & Value\\
    \hline
    Time scale & $T$ & \SI{1e-2}{s}\\
    Length scale & $L$ & \SI{1e-4}{m}\\
    Surface tension scale & $\sigma_0$ & \SI{1e-2}{kg.s^{-2}}\\
    Mass density scale & $\rho_0$ & \SI{1e3}{kg.m^{-3}}\\
    Mass density variation scale & $\Delta \rho_0$ & 0, \SI{1e3}{kg.m^{-3}}\\
    Kinematic viscosity & $\nu_0$ & \SI{1e-6}{m^2.s^{-1}}\\
	Gravity & $g$ & \SI{1e2}{m.s^{-2}}\\
	\hline
	\end{tabular}
	\caption{Dimensional parameters for simulations.}
	\label{tab:dimensional-parameters}

	\vspace{1em}	
	\begin{tabular}{lcl}
    Name & Symbol & Value\\
    \hline
    Interface thickness & $\eps$ & \num{1e-2}\\
    Bond number & Bo & \num{1e-3}\\
    Capillary number & Ca & \num{1e-3}\\
    Weber number & We & \num{1e-3}\\
    Surface tensions & $\sigma_{1,2}, \sigma_{1,3},\sigma_{2,3}$ & $0.0636,	0.4983,	0.4381$\\
    Density perturbations & $d\rho_1, d\rho_2,d\rho_3$ & $0.05,-0.01,0.6$\\
    Initial concentrations & $c_{1,0}, c_{2,0}$ & $0.3,0.4,\ldots,0.8$\\
    Noise magnitude & $\Delta c$ & 0.05\\
	\hline
	\end{tabular}
	\caption{Nondimensional parameters for simulations.}
	\label{tab:nondimensional-parameters}

	\vspace{1em}	
	\begin{tabular}{ll}
    Quantity & Value\\
    \hline
    Spatial modes $n_x,n_y,n_z$ & 192, 192, 384\\
    Dealias factor & 3/2\\
    Time step $dt$ & \num{1e-4},\num{2e-4} \\
    Time step scheme & SBDF2\\
	\hline
	\end{tabular}
	\caption{Numerical parameters for simulations.}
	\label{tab:numerical-parameters}
\end{table}

\subsection{Initial conditions}
Mixed initial conditions of $c_1, c_2$ are given by balls of radius $0.7$ that smoothly transition from 0 outside the ball to $\{c_{1,0}, 1-c_{1,0}\}$ inside the ball, which are perturbed within the ball by uniformly distributed random Fourier and Chebyshev coefficients between $-\Delta c/2$ to $\Delta c/2$, and where behaviour near the boundary is given by a tanh profile of thickness $\sqrt{2} \eps$:
    \begin{align}
    c_1(0,x,y,z) &= K(r(x,y,z)-0.7)(c_{1,0} + \Delta c \text{ Noise}),\\
    c_2(0,x,y,z) &= K(r(x,y,z)-0.7)((1-c_{1,0}) - \Delta c \text{ Noise}),\\
    K(q) &= \frac{1}{2}\br{1 + \tanh\br{-\frac{q}{\sqrt{2}\eps}}},\\
    r &= \sqrt{x^2 + y^2 + z^2},\\
    \text{Noise} &= \sum_{j=0}^{64} \sum_{k=0}^{64} \sum_{\ell=0}^{128} C_{i,j,k} e^{i \pi j x} e^{i \pi ky } T_\ell(2z - 1) + \text{c.c.},\\
    C_{i,j,k} &\sim \mathcal{U}(-\tfrac{1}{2}\Delta c, \tfrac{1}{2}\Delta c).
    \end{align}
The 2D simulations simply take the zero $y$ mode as an initial condition.
All mixed simulations use the same random noise coefficients.
For the separated initial conditions in 2D, the initial conditions are instead given by
    \begin{align}
    c_1(0,x,z) &= K(r(x,z)-0.7) K(z),\\
    c_2(0,x,z) &= K(r(x,z)-0.7) K(-z).
    \end{align}

\section{Simulation analysis}
\renewcommand{\thefootnote}{\arabic{footnote}}
We quantitatively analyse each 2D simulation using python as follows.
\begin{enumerate}
    \item For each simulation we run through each save. For each save, we:
    \begin{enumerate}
        \item Reinterpolate onto a regular uniform grid with a second order spline\footnote{using \texttt{scipy.interpolate.RectBivariateSpline} from scipy at \href{https://docs.scipy.org/doc/scipy/reference/interpolate.html}{https:/docs.scipy.org/doc/scipy/reference/interpolate.html}}.
    
        \item Recentre the simulation in the horizontal direction by calculating the grid point furthest from the $c_3$ phase boundary using the fast marching method\footnote{using \texttt{skfmm.distance} from scikit-fmm at \href{https://github.com/scikit-fmm/scikit-fmm}{https://github.com/scikit-fmm/scikit-fmm}}, and shifting it to $x=0$.

        \item Isolate the $c_1$ and $c_2$ droplets by taking binary thresholds above 0.7.

        \item Label and count individual droplets using feature detection algorithms from scipy.ndimage\footnote{using \texttt{scipy.ndimage.label} from \texttt{scipy.ndimage} at \href{https://docs.scipy.org/doc/scipy/reference/ndimage.html}{https://docs.scipy.org/doc/scipy/reference/ndimage.html}.}. Horizontal, vertical and diagonal connections are considered adjacent\footnote{i.e. \texttt{structure=np.ones([3,3])} in 2D and \texttt{structure=np.ones([3,3,3])} in 3D.}.

        \item Classify the configuration of the snapshot as
        \begin{itemize}
            \item \textbf{Nonequilibrium} if there are two or more droplets of either phase, otherwise:
            \item \textbf{${c_2}$ drop} if there are no $c_1$ droplets, otherwise:
            \item \textbf{${c_1}$ drop} if there are no $c_2$ droplets, otherwise:
            \item \textbf{Crescent} if at any point $c_1 >0.3$ and $c_3 > 0.3$, otherwise:
            \item \textbf{Shell} if $\max c_1 > 0.9$ and $\max c_2 > 0.9$.
        \end{itemize}

    \end{enumerate}

    \item For each simulation, determine the first time at which the configuration has the same value as the final configuration.
\end{enumerate}

\begin{figure}[h]
    \centering
    \includegraphics[width=\linewidth]{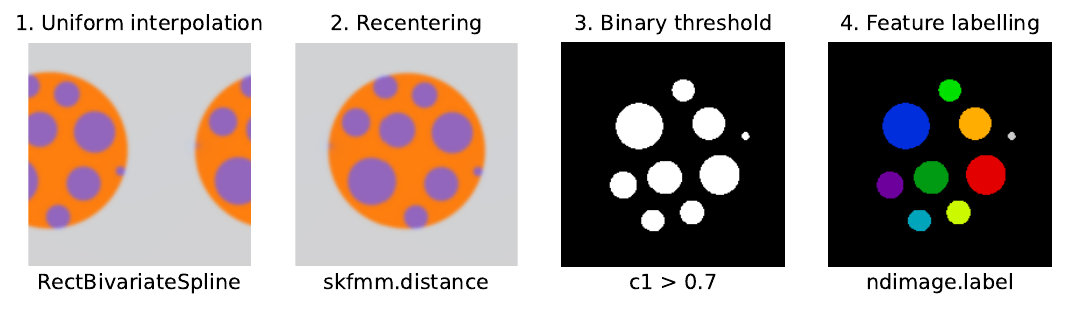}
    \caption{
    Overview of analysis pipeline for each time step for the 2D simulations.
    }
    \label{fig:particle-number-vs-time}
\end{figure}

\section{How does $\varepsilon$ affect coarsening?}
\begin{figure}[h]
    \centering
    \includegraphics[width=.7\linewidth]{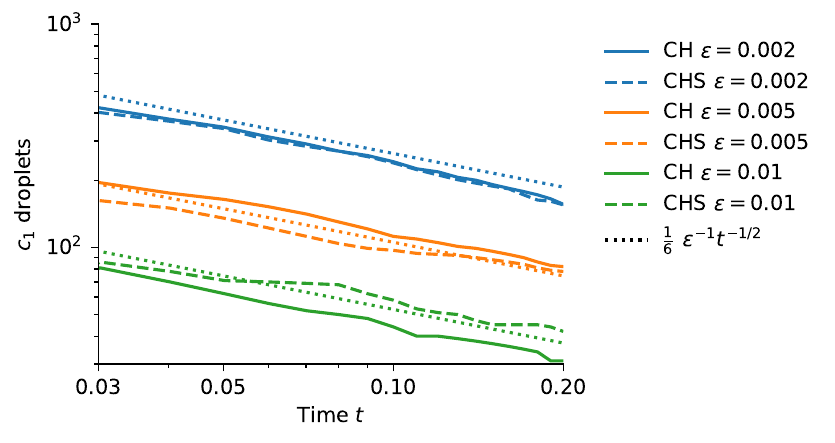}
    \caption{
    Dependence of coarsening time on interfacial thickness parameter $\eps$.
    The number of particles at time $t$ is approximately proportional to $\eps^{-1}t^{-1/2}$.
    We therefore expect the time to complete coarsening to scale as $\eps^2$.
    }
    \label{fig:droplet-number-eps}
\end{figure}

The interfacial thickness parameter $\eps$ determines not only the thickness of the diffuse interface between phases, but also the separation of time scales.
Comparing three different choices of $\eps$ (\num{1e-2}, \num{5e-3}, \num{2e-3}) with initial concentration $c_{1,0} = 0.3$, for the same initial conditions as in section 4 A, we find that the number of $c_1$ droplets over time is proportional to $\eps^{-1}t^{-1/2}$ (\cref{fig:droplet-number-eps}). Note that the spatial resolution ($\varepsilon\, n_x$ and $\varepsilon \,n_y$) in each case is constant; see table \ref{tab:eps-numerical-parameters}. 
This accords with figure 7 in the main paper, where we find that the number of droplets is approximately proportional to $t^{-1/2}$.

Writing the functional dependence of droplet number on $t$ and $\eps$ as
    \begin{align*}
    \text{\# droplets} \equiv f(\eps, t) \propto \eps^{-1}t^{-1/2},
    \end{align*}
we fit this function to the simulation data in \cref{fig:droplet-number-eps}, and find an approximate constant of proportionality $f(\eps, t) \approx \frac{1}{6}\eps^{-1}t^{-1/2}$. The accuracy of the fit generally improves with increasing $t$ and decreasing $\varepsilon$.
Therefore we can predict the time to complete coarsening as 
    \begin{align*}
    f(\eps,t) = 1 \implies t &= 36\eps^2.
    \end{align*}
The ratio of experimental to simulation time scales ($\approx 10^5$), tells us the physical interfacial thickness is up to $10^{-5/2}$ smaller than the simulation thickness, corresponding to \SI{1e-8}{m} rather than $\eps L = \SI{1e-6}{m}$. This $\mathcal{O}(\SI{1}{nm})$ interfacial thickness is consistent with previous measurements \cite{MagalettiSharpinterfaceLimitCahn2013}. 
We note the fit of the scaling law improves as $\eps$ decreases, as would be expected with larger numbers of droplets.
The constant of proportionality may change in three dimensions, but we do not expect a change in scaling law.

\begin{table}[h!]
    \centering
    \begin{tabular}{ll}
    Quantity & Value\\
    \hline
    Model & CH, CHS\\
    Initial concentration $c_{1,0}$ & 0.3\\
    Interfacial thickness $\eps$ & $$\num{1e-2}, \num{5e-3}, \num{2e-3}$$\\
    Time step $dt$ & $$\num{2e-4}, \num{2e-4}, \num{5e-4}$$\\
    Time step scheme & SBDF2\\
    Fourier modes $n_x$ & 192, 384, 512\\
    Chebyshev modes $n_z$ & 384, 768, 1024\\
    Dealias factor & 3/2\\
    Processors & 8\\
    \hline
    \end{tabular}
    \caption{Numerical parameters for varying $\eps$ simulations.}
    \label{tab:eps-numerical-parameters}
\end{table}

\renewcommand{\thefootnote}{\fnsymbol{footnote}}

\FloatBarrier
\section{Movies and Datasets}

\movie{2D-simulations.mp4}

Side-by-side movie of 2D CH, CHS, CHSB $z$, and CHSB $x$ models for varying initial concentration ratios.\\

\movie{3D-simulations.mp4}

Side-by-side movie of 3D CH, CHS, CHSB $z$, and CHSB $x$ models.\\

\movie{Experiment-PEG33-Gel66.mp4}

Movie of temperature-induced phase-separation of a PEG-gelatin aqueous two phase system. Experimental details are listed in the methods section of the main paper.\\

All code and parameters required to generate data are provided at \href{https://github.com/ericwhester/multiphase-fluids-code}{github.com/ericwhester/multiphase-fluids-code}.